# From Critical Transitions to Sustainability Transformations: Phenomena, Theory, and Applications Across Disciplines

**Authors:** E.H. Krueger[1*], K. Halder[2], W. Oh[3], N. Molla[4,5], J.C. Rocha[5], G. Ferro[4], G. Falmagne[6], A. B. Stephenson[4,17], L. Wang Erlandsson[5,7,8], J. Bara[9,10], G. I. Hagstrom[11], N. Wunderling[7,12,13], J. Li[4,18]; X. Huggins[4,5,14,15], E.R. Zajdela[4,15], H. Jaggi[4], S. M. Constantino[19], C. Schill[4,20,21], V.V. Vasconcelos[10,22]

**Author Affiliations:** 1: Institute for Biodiversity and Ecosystem Dynamics (IBED), University of Amsterdam, The Netherlands; 2: Department of Ecology, Evolution & Natural Resources, Rutgers University, USA; 3: Department of Systems Design Engineering, University of Waterloo, Canada; 4: High Meadows Environmental Institute, Princeton University, USA; 5: Stockholm Resilience Centre, Stockholm University, Sweden; 6: Technological University Dublin & Trinity College Dublin, Ireland; 7: Potsdam Institute for Climate Impact Research (PIK), Germany; 8: Bolin Centre for Climate Research, Stockholm University, Sweden; 9: University of Bonn, Germany; 10: Computational Science Lab, Institute for Informatics, University of Amsterdam, The Netherlands; 11: City University of New York, USA; 12: Center for Critical Computational Studies, Goethe-University Frankfurt/Main, Germany; 13: Senckenberg Research Institute and Natural History Museum, Frankfurt/Main, Germany; 14: Institute for Resources, Environment and Sustainability, University of British Columbia, Canada; 15: School of Geographical and Earth Sciences, University of Glasgow, Scotland; 16: Santa Fe Institute, USA; 17: Alignment of Complex Systems Research Group, Prague, Czech Republic; 18: Program in Applied and Computational Mathematics, Princeton University, USA; 19: Stanford Doerr School of Sustainability, Stanford University, USA; 20: Beijer Institute of Ecological Economics; Anthropocene Laboratory, Royal Swedish Academy of Sciences, Sweden; 21: Anthropocene Laboratory, Royal Swedish Academy of Sciences, Stockholm, Sweden; POLDER Centre, Institute for Advanced Study, University of Amsterdam, The Netherlands.

**Corresponding author:** E.H.Krueger: e.h.krueger@uva.nl; phone: +31(6)21481864; ORCID: 0000-0003-0630-3363

**Author emails:** K. Halder: kh832@sebs.rutgers.edu; W.Oh: woisok.oh@uwaterloo.ca; N. Molla: nusrat.molla@princeton.edu; J.C. Rocha: juan.rocha@su.se; G. Ferro: giuseppe.ferro@princeton.edu; G. Falmagne: Guillaume.Falmagne@TUDublin.ie; A.B. Stephenson: stephenson@princeton.edu; L. Wang-Erlandsson: lan.wang@su.se; J. Bara: j.bara@uva.nl; G.I.Hagstrom: gih@princeton.edu; N. Wunderling: nico.wunderling@pik-potsdam.de; J. Li: jl2894@princeton.edu; X. Huggins: Xander.Huggins@glasgow.ac.uk; E.R.Zajdela: emma.zajdela@princeton.edu; H.Jaggi: harmanj@princeton.edu; S.M.Constantino: sara.constantino@gmail.com; C.Schill: caroline.schill@beijer.kva.se; V.V.Vasconcelos: v.v.vasconcelos@uva.nl

**Short title:** Critical Transitions across disciplines

## Abstract

There has been a recent surge in the use of terms describing systemic change processes across science, policy, and public debate. The associated change phenomena are often assumed to be abrupt and irreversible once critical thresholds are crossed. However, the interchangeable use of terms leaves definitions and the underpinning scientific theory unclear, which can lead to misconceptions about transition processes and potential impacts. Here, we present selected empirical examples representing phenomena of tipping points, regime shifts, critical transitions, phase transitions, social and socio-technical transitions, and sustainability transformations, together with their underlying theory derived from the broadly defined domains of physics and mathematics, Earth science and ecology, social sciences, and their interdisciplinary overlaps. We bring these fields into closer conversation, creating a shared understanding of each of the conceptual roots, phenomena, mechanisms, and specific applications. This can help steward the governance of social-ecological change towards more just and sustainable futures.

## Significance Statement

We increasingly frame sustainability goals around tipping points and transformations: some to avoid, such as Amazon dieback, others to trigger, such as the switch to clean energy. The underlying ideas have travelled in several directions: from critical and phase transitions in physics and mathematics into regime shifts in ecology, but also from sociology, where "tipping point" was coined, into climate science, and have both lost precision and gained applicability on the way.

Reviewing eleven cases, from magnets and epidemics to lakes, segregated neighbourhoods and energy systems, we show which properties (abruptness, irreversibility, early warning) carry over between fields and which do not. The result is a set of questions for judging whether a claimed tipping point is real, measurable and steerable, and for whom.

## 1. Introduction

Tipping points, regime shifts, critical transitions, phase transitions, social and socio-technical transitions, and sustainability transformations describe a variety of scientifically and societally relevant biophysical and social change processes. The associated empirical phenomena covered by these terms range from the rapid spread of epidemic diseases (1), the shift from coral to algal dominance in reefs (2), to long-lasting changes in Earth's climate (3). In society, the 'tipping' of opinions between political extremes or social norm dynamics have led to widespread changes, which range from the shift from smoking to non-smoking as the norm (4) to major socio-political disruptions, such as the end of South Africa's Apartheid in the early 1990s (5). Some formal models describing these phenomena share their mathematics with transitions in physics, such as the freezing and melting of water or materials buckling under pressure; others arose independently in ecology and the social sciences, and ideas have travelled in both directions.

The recent surge in the use of the mentioned terms across disciplines and in policy and public debate focuses on either feared, fast and persistent change (e.g., climate change) - or on needed change in one domain (e.g., behavioural change) to avoid feared change in another (6). The terms used to describe these systemic changes are often used loosely and interchangeably, while definitions are often unclear. This can lead to misconceptions about their underlying processes and resulting impacts, and the potential dynamics that enable or hinder positive societal change (7–9).

Recent research in the closely related fields of 'critical transitions' and 'phase transitions' in physics, has reconciled mismatches between expected and observed transition characteristics by extending the theory to include other physical mechanisms and types of transitions into their mathematical description (10). Concerns about the use of the term 'tipping point' (9) have also prompted cross-system definitions that emphasize specific properties and mechanisms (11). Similarly, a review of the conceptual development of 'regime shifts' and 'transformations' research shows the need for stronger integration and cross-fertilization to promote more powerful approaches to navigating transitions (12).

In this review, we extend this integration across a broader set of transition phenomena and scientific fields. We review transition phenomena, their underlying theory, and its application across domains, with the purpose of making transparent the roots of these models, how they are applied across fields, and the limitations of capturing phenomena in scientific theory. We inquire how concepts and theories have travelled across scientific domains and how further cross-fertilization could advance our ability to analyze these phenomena. We open a closer conversation between the research fields addressing societal and biophysical transformations, to strengthen the understanding of the mechanisms and dynamics of change processes and bridge their analytic tools across systems; this can help steward the governance of changing environmental and social systems (13).

To start this cross-field conversation, we extract literature covering six canonical terms - tipping points, regime shifts, critical transitions, phase transitions, social and socio-technical transitions, and sustainability transformations - jointly referred to under the term 'transitions'. The topics covered by the almost 90,000 articles, published between 1870-2024, show that, for phase transitions (>60,000 articles), classical topics, such as magnetism and applications in astronomy have declined in relative frequency, while quantum topology, black holes, and material chemistry grew over time. For the other terms (>20,000 articles), topics related to the collapse of fisheries in marine ecosystems, financial markets, the physics of superconductors and quantum studies dominated the transitions literature in the early 2000's and have declined since then. Instead, topics related to bifurcation theory and time series, urban and rural studies, sustainability transformations, and regime shifts in terrestrial ecosystems gained attention (detailed topic modeling results are presented in the Supplementary Information, SI). The increased attention to societally relevant topics reveals how critical transition vocabulary is being used to describe phenomena across fields.

In what follows, we present selected empirical phenomena and characteristics of transition processes associated with conceptualizations of each of the six terms, together with their underlying theory. These examples cover four domains (Figure 1): physics and mathematics, Earth systems and ecology, the social sciences, and sustainability transformations, and include interdisciplinary examples that intersect multiple domains. We then conduct a co-citation network analysis, which illustrates how terms, concepts, and theories have migrated across different research fields over time. We conclude by proposing potential future avenues for research that benefit from these insights across the fields.

## 2. Transition phenomena and expanding theoretical frameworks

The examples presented here include classical, recent, and interdisciplinary examples, which have made significant contributions to the development of the respective transitions field. For each example, we describe the phenomenon and its defining transition characteristics together with the theory, model(s), or frameworks that are used to generalize the mechanisms and dynamics of the transition phenomena. Where applicable, we discuss interdisciplinary applications of the models and the current research frontier for each domain, whether this is in the further development of the theory, estimations of risks and consequences of the described transitions, or challenges and insights for their management/governance. The selection of examples and the theory used to describe them is based on the salience and breadth of the examples, as well as the expertise of the contributing authors.

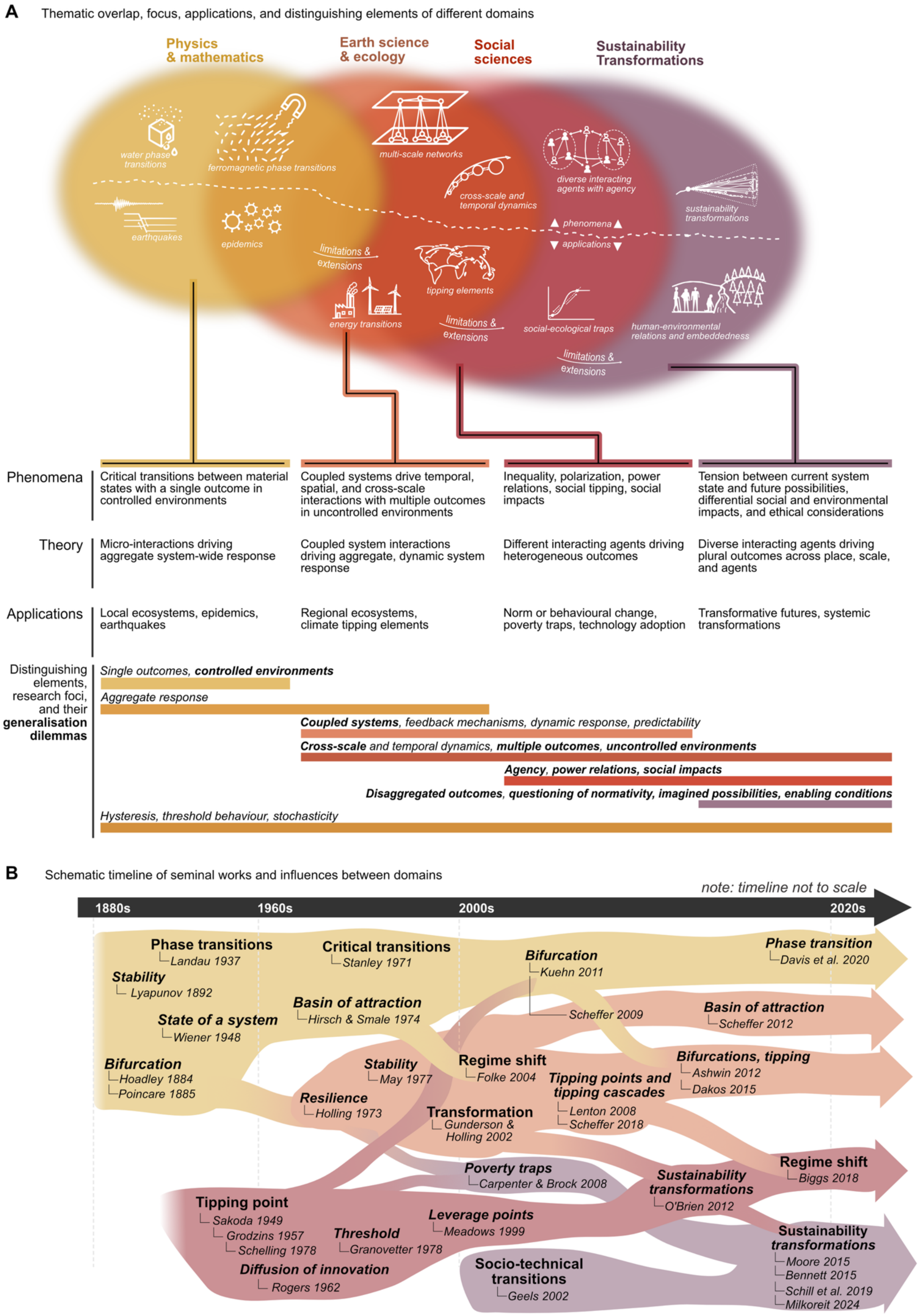
A
Thematic overlap, focus, applications, and distinguishing elements of different domains
Physics & mathematics
Earth science & ecology
Social sciences
Sustainability Transformations
water phase transitions
ferromagnetic phase transitions
multi-scale networks
diverse interacting agents with agency
cross-scale and temporal dynamics
sustainability transformations
earthquakes
epidemics
phenomena
applications
limitations & extensions
tipping elements
energy transitions
limitations & extensions
social-ecological traps
human-environmental relations and embeddedness
limitations & extensions
Phenomena
Critical transitions between material states with a single outcome in controlled environments
Coupled systems drive temporal, spatial, and cross-scale interactions with multiple outcomes in uncontrolled environments
Inequality, polarization, power relations, social tipping, social impacts
Tension between current system state and future possibilities, differential social and environmental impacts, and ethical considerations
Theory
Micro-interactions driving aggregate system-wide response
Coupled system interactions driving aggregate, dynamic system response
Different interacting agents driving heterogeneous outcomes
Diverse interacting agents driving plural outcomes across place, scale, and agents
Applications
Local ecosystems, epidemics, earthquakes
Regional ecosystems, climate tipping elements
Norm or behavioural change, poverty traps, technology adoption
Transformative futures, systemic transformations
Distinguishing elements, research foci, and their generalisation dilemmas
Single outcomes, controlled environments
Aggregate response
Coupled systems, feedback mechanisms, dynamic response, predictability
Cross-scale and temporal dynamics, multiple outcomes, uncontrolled environments
Agency, power relations, social impacts
Disaggregated outcomes, questioning of normativity, imagined possibilities, enabling conditions
Hysteresis, threshold behaviour, stochasticity
B
Schematic timeline of seminal works and influences between domains
note: timeline not to scale
1880s
1960s
2000s
2020s
Phase transitions
Landau 1937
Critical transitions
Stanley 1971
Bifurcation
Kuehn 2011
Phase transition
Davis et al. 2020
Stability
Lyapunov 1892
Basin of attraction
Hirsch & Smale 1974
Scheffer 2009
Basin of attraction
Scheffer 2012
State of a system
Wiener 1948
Bifurcation
Hoadley 1884
Poincare 1885
Stability
May 1977
Regime shift
Folke 2004
Tipping points and tipping cascades
Lenton 2008
Scheffer 2018
Bifurcations, tipping
Ashwin 2012
Dakos 2015
Resilience
Holling 1973
Transformation
Gunderson & Holling 2002
Poverty traps
Carpenter & Brock 2008
Sustainability transformations
O'Brien 2012
Regime shift
Biggs 2018
Tipping point
Sakoda 1949
Grodzins 1957
Schelling 1978
Threshold
Granovetter 1978
Leverage points
Meadows 1999
Diffusion of innovation
Rogers 1962
Socio-technical transitions
Geels 2002
Sustainability transformations
Moore 2015
Bennett 2015
Schill et al. 2019
Milkoreit 2024

**Figure 1: A | Phenomena, theory, and applications of 'transitions'** across the four domains with distinguishing elements and research foci in each. Models and theory developed in one domain are limited in representing particular phenomena observed in others, which leads to extensions, adaptations, and novel developments. **B | Schematic timeline** with seminal works marking the development of transitions theory across the four domains.

## 2.1 Roots in physics and mathematics: Networked transitions, collective behavior, and criticality

The three examples, two from physics and one from epidemiology, a field with its own threshold mathematics, provide the vocabulary and models that several other fields adopt. They distinguish "critical" transitions, which are often continuous, from changes that are simply abrupt, large, or fast. They provide the foundational mechanistic description of how interactions among microscopic elements can produce qualitative changes at macroscopic scales, and they underpin the early warning indicators now used across fields. They also show how network structure can generate cascades or make an expected threshold disappear. Because similar dynamics arise in social and ecological systems, these insights are also a cautionary tale of how thresholds are sought in social and ecological systems.

### 2.1.1 Phase transitions in physics: Ferromagnetism and water

#### The phenomenon & the model

Certain metals become magnetic when cooled below a critical temperature. This is the ferromagnetic phase transition, well described by the Ising model. In this model, magnetic moments at the (sub)atomic scale, called spins, can point either up or down. The macroscopic state of the system is characterized by its net magnetization, the sum of all microscopic magnetic moments. The model captures the material's macroscopic behavior by considering the competition between local interactions among neighboring spins, which tend to align their directions, and temperature, which randomizes them. At high temperature, thermal fluctuations dominate over the strength of local interactions, so the spins remain disordered and there is no net magnetization. As temperature (the control parameter) decreases toward a critical value, the Curie temperature, local alignment becomes increasingly coordinated across larger spatial regions. At the critical temperature, correlations extend over very large distances throughout the system, despite the interactions themselves being only local, and below this temperature a macroscopic magnetization appears (a qualitative change in the macroscopic state, the order parameter). Thus, the system transitions from a non-magnetized phase into a magnetized phase. This transition illustrates how simple local interactions can generate collective order once a precise threshold is crossed (14).

### Critical transition characteristics

Transitions such as the ferromagnetic one are termed "critical" because near the critical point (the control-parameter value at which they occur), fluctuations in magnetization occur across a very broad range of spatial scales, the correlation length becomes very large, and the system becomes highly susceptible to external perturbations. Ferromagnetic transitions are continuous, or second-order, phase transitions: The order parameter changes continuously, while the response quantity, such as the magnetic susceptibility, can change abruptly at the critical point (15). This transition displays symmetry-breaking: in the disordered phase there is no preferred direction of magnetization, whereas in the ordered phase the system spontaneously selects one. It also exhibits universality: near the critical point, the behavior of the order parameter, correlation length, and susceptibility, follows characteristic power law distributions. These power laws, described by critical exponents that represent the system's scaling behavior near criticality, unite many otherwise very different physical systems into universality classes, even when their microscopic components and detailed interactions differ. Because changes in the dynamics of fluctuations, correlations, or susceptibility can sometimes be detected as the critical point is approached, such behavior might signal an approaching transition. Hysteresis (where the transition point depends on the direction of change) is a term coined for magnetized iron (16). It occurs when a ferromagnet below the Curie temperature is driven by an external field: magnetization flips at different field intensities on the increase and decrease, which is how magnets store information, capturing the history of the system.

Other phase transitions, such as water changing from solid to liquid to vapor, are first-order, in which the order parameter changes discontinuously at the transition point. Here, energy ('latent heat') is absorbed or released, while the temperature remains constant, and macroscopic properties such as density, volume, or entropy can change abruptly. These transitions can be fully reversible: reversing the energy exchange can return the system to its previous phase. Thus, the term "critical" in physics refers primarily to collective reorganization, long-range correlations, and strong sensitivity near a transition, rather than to abrupt regime shifts or irreversibility, as the term can sometimes imply in other fields. The term 'critical point' (17) marks the point where the jump between liquid and gas vanishes. Thus, critical transitions do not necessarily involve abrupt jumps in macroscopic state variables. Importantly, features of criticality, such as extremely large correlation lengths and susceptibilities are exact mathematical properties only in the thermodynamic limit, that is, for systems containing an effectively infinite number of microscopic elements.

### 2.1.2 Collective failure in materials: cascades and instability

#### The phenomenon and its characteristics

The catastrophic failure of materials under stress provides a striking example of critical transitions: in materials represented as a large bundle of interconnected fibers subjected to a gradually increasing load, each intact fiber supports part of the total stress and has a characteristic threshold beyond which it breaks. When these thresholds vary across fibers, increasing stress can trigger failure cascades: once one fiber breaks, the stress is redistributed among the remaining intact fibers, causing some of them to exceed their own thresholds and fail in turn. Cascades of this kind can eventually destroy a large fraction of the bonds that originally sustained the material (18–20).

The nature of this failure depends on the material's microscopic properties, including the distribution of failure thresholds and the way stress is redistributed after individual fibers break. If the fibers are homogeneous, the failure of the first few elements can trigger an abrupt collapse of the system, without a preceding cascade. If the distribution of failure thresholds is broad, damage may accumulate gradually with increasing stress, because stronger fibers can carry the additional load released by weaker ones, meaning no single catastrophic threshold can be observed. For intermediate levels of heterogeneity, the system can show a clear form of critical behavior: failure cascades become progressively larger as global collapse approaches (21, 22).

This example illustrates the difference between criticality and abruptness. The final macroscopic failure of the material may be sudden, but the dynamics leading up to that failure can nevertheless show characteristics of a critical transition. The sizes of failure cascades can follow power-law distributions, described by universal critical exponents, with frequent small events and large events becoming increasingly likely as the system approaches breakdown.

Scale-free cascade dynamics can signal an approaching transition, when perturbations propagate and cascade through increasingly large parts of the system (10, 23). Spatial organization can provide additional signatures of criticality: cracks may develop complex, scale-dependent structures characterized by non-trivial fractal dimensions (24, 25). Additionally, these materials can also exhibit memory and path-dependence, with their response (such as acoustic emissions from fractures) depending on the system's loading history (26).

### The model

In the language of statistical physics, these failure transitions can be understood as spinodal instabilities. A spinodal instability occurs when a metastable state—that is, a state that is locally stable but not globally optimal—progressively loses stability as a control parameter is varied. At the spinodal point, this metastable state disappears, forcing the system to reorganize rapidly into a different state (25, 27). This framework describes catastrophic transitions in several nonequilibrium systems. Models such as the fiber bundle model can display spinodal-like instabilities and have been used to study not only material fracture but also other collective failure phenomena, including earthquakes (28).

### Interdisciplinary applicability of the model

Many social and environmental systems share the features of these models: many heterogeneous components whose states depend on others (10, 29). Granovetter's threshold model (Section 2.3.2) has the same structure as the fiber bundle model, and the Random Field Ising Model has been used to model opinion shifts, social contagion, and financial instability (30–33), with complex interaction networks shaping cascades (34).

In ecology, phase-separation models (the physics of spinodal decomposition, distinct from the spinodal instability above) explain self-organized spatial patterns and their resilience (35, 36).

Although the mechanisms and meanings of the relevant variables differ across materials, ecosystems, and people, the underlying models share important mathematical features: heterogeneous local thresholds, interactions among components, metastable states, cascading responses, and the loss of stability as a control parameter changes.

### 2.1.3 Epidemics: Connectivity, percolation, and network structure

#### The phenomenon

The human-to-human or animal-to-human transmission of pathogens that cause infectious diseases, such as measles, the flu, or Covid-19, are a central focus of epidemiology. These epidemics often occur in waves, which are driven by the dynamics between infected, susceptible and immune individuals (1, 37). As new diseases and new variants of existing ones emerge, epidemics may break out, causing an infection-free population to transition to a state of widespread disease.

#### The model

In the classic Susceptible-Infected-Recovered (SIR) model used to study disease transmission, susceptible (S) individuals get infected (I) with a certain probability when interacting with an infected individual. Infected individuals may then recover (R) at a certain rate (38). In the simpler Susceptible-Infected-Susceptible (SIS) model, where individuals do not gain immunity after having been infected, the transition to the endemic state, where the disease is sustained in the population, is formally a transcritical bifurcation (i.e., where two equilibria approach each other). The transition is determined by the basic reproduction number, which characterizes the average number of secondary infections by one infectious individual in a completely susceptible population. That is, a reproductive number below the critical threshold of 1 leads to an infection-free state, while a reproductive number above the threshold leads to an exponential increase of the infection until the endemic state. For any given disease, this number depends on biological and physiological characteristics of the disease and the carrier, as well as the social properties of the population.

### Another model

Beyond this direct formalisation within bifurcation theory (see below), epidemic spread has been connected to percolation theory and the structure of social networks: the final size of an SIR outbreak maps onto bond percolation (39), which can also be modeled on networks (40). Here, for a small fraction of infected individuals who infect others with a low probability, only small clusters of infection exist; as probability grows, clusters grow and merge, and above a critical value a system-spanning component appears. At this threshold, the size distribution of clusters follows a power law, and the correlation length becomes very large, indicating scale invariance—as is the case in the ferromagnetic transition detailed earlier. The transition does not involve energy minimization (where the stable states at each end of the transition are those with the least energy) or symmetry breaking; instead, it marks the emergence of large-scale connectivity from random local structure. As the SIR model is formally equivalent to this percolation transition, the epidemic threshold below which the disease cannot spread is linked to the percolation threshold of the network that the population sits on.

### Empirical attempts derived from the models

The connection to network percolation has been the ground for rich investigation of epidemics on networks, which shows that, in many types of network structures, the epidemic threshold vanishes in large networks, so any infection rate can sustain the disease (41–43).

Models of epidemic spread also adopt techniques from spatial pattern formation, such as cellular automata and ecological spatial models. Here, models of disease transmission create spatial infection patterns that transition from self-limiting local outbreaks to turbulent waves traveling across space, and stable spiral waves as infection rate increases, with the system shifting from disease extinction to disease persistence (44). Those spatial/temporal patterns predicted by the theoretical model are observed in real world epidemiological systems (45–47). The emergence of such patterns is sometimes considered as an early warning signal of disease outbreak (48).

#### 2.1.4 Mathematical foundations: bifurcations, stochasticity, and tipping mechanisms

When described as dynamical systems, several of these phenomena correspond to bifurcations; the abrupt ones, such as the spinodal failure of materials, are examples of B-tipping (bifurcation-induced tipping), which reoccur below in the examples from the Earth sciences and ecology. They include fold (or saddle-node) bifurcations, commonly used to describe critical transitions in life and Earth sciences (49). In the context of B-tipping, the term 'critical' denotes the control-parameter value at which an equilibrium loses stability (mathematically, the eigenvalue of the Jacobian reaches zero real part) (49, 50). This mathematical bifurcation theory studies the structure and stability of equilibria in dynamical systems (51–53). It relies on "fast-slow" systems (50), where the state variable (the order parameter) relaxes fast and closely follows the equilibrium because the control variable changes slowly enough.

At the macroscopic level, the statistical physics and tipping literatures describe transitions with the same bifurcation theory but use 'critical' for different things. In physics, a critical point marks a continuous transition; in the tipping literature, a critical transition is an abrupt jump (49, 50). Both involve a loss of stability, but the type of bifurcation differs (54): in 'safe' bifurcations (e.g. supercritical pitchfork or Hopf) the system moves smoothly onto a nearby new state; in 'dangerous' ones (e.g. fold or subcritical Hopf) no nearby state remains and the system jumps. In mean-field models, a spinodal is a fold and the critical point is the cusp where two folds meet.

Definitions of 'tipping point' also differ in their required properties: Lenton et al. (55) admit gradual change, whereas Abrams et al. (11) require feedback, a threshold, and persistence. Abruptness alone is neither necessary nor sufficient. Ashwin et al. (49) classify tipping by its cause: bifurcation-induced tipping (B-tipping), when a slowly varied parameter crosses a dangerous bifurcation; noise-induced tipping (N-tipping), when fluctuations push the system out of its attractor's basin; and rate-induced tipping (R-tipping), when a parameter changes too fast for the system to track its moving attractor. Further processes have been proposed to describe transitions, including phase- (56) and shock-induced tipping (57).

Categorizing real-world transitions depends on the modeling framework and assumptions, including the choice of state and control variables and the timescale used to define equilibria (58). Because social and ecological systems are high-dimensional, reducing them to low-dimensional models can miss dynamics that matter for the classification.

### 2.2 Ecological regime shifts and Earth-system tipping elements: feedbacks, environmental context, and scale

The field of Earth Science and Ecology has focused on leveraging mathematical and computational models for understanding the mechanisms that lead to spatial and temporal tipping dynamics and on developing and measuring resilience indicators and early warning signals. Our examples cover different scales of transitions from lake eutrophication to forest-savannah-grassland transitions and Earth system tipping elements across which confidence in the mechanisms holds while confidence in the thresholds’ existence and values weakens. At the same time, the socio-economic costs associated with ecological transitions make managing these systems to avoid regime shifts particularly salient.

### 2.2.1 Lake eutrophication: alternative regimes, feedbacks, and reversibility

#### The Phenomenon

Shallow lakes are among the best-documented ecosystems posited to occupy two alternative stable states: a clear-water state dominated by submerged plants, and a turbid state dominated by phytoplankton (59). The persistence of both states is attributed to self-reinforcing feedbacks: Abundant aquatic plants tend to perpetuate a clear state by inhibiting phytoplankton development and sediment re-suspension, whereas light limitation in turbid states prevent the establishment of aquatic plants that can help re-instate the clear state. The strength and existence of alternative stable state dynamics depends on a range of factors, such as lake depth profile (60), spatial heterogeneity, the degree of water circulation (61), and fluctuation of environmental conditions (62). Other studies show that the relationship between nutrients and phytoplankton biomass remains linear at a multi-year time scale, without exhibiting alternative stable state dynamics (63). In some lakes, regular oscillations between clear and turbid states have been observed and attributed to an interplay between fast and slow dynamics: i.e., while submerged plants establish clear conditions in the short term, they also enhance nutrient accumulation that eventually leads to phosphorous release from sediments that punctuates their dominant position (64). Further, different types of alternative stable states and regime shifts have been suggested beyond the original proposal of the clear and turbid states, due to the ability of different plant species, algae, or cyanobacteria to stabilize their dominance (65, 66).

### Theories, models, experiments

Lakes became the classical empirical case for the theory of alternative stable states introduced by Holling and May (67, 68), and were proposed to exhibit alternative stable states based on a simple conceptual vegetation-algae model (59). Lakes served as the main test bed for statistical early warning signals. Rising variance, increasing autocorrelation, and low return rates anticipated lake transitions more than a year before a manipulated food web completed its shift in a whole-lake experiment (69). In another, halting nutrient inputs once indicators crossed preset bounds reversed a developing cyanobacterial bloom (70). However, while indicators of approaching transitions in five well-documented shifts in natural lakes were often detected, sometimes years ahead, these indicators agreed poorly with one another (71). The term ‘alternative regime’ has been proposed in lieu of ‘alternative state’ to capture the fact that lakes are in dynamic, fluctuating regimes rather than in stable states (62).

### Risks, management

Eutrophied lakes with anoxic conditions lose the ability to provide important ecosystem services for, e.g., fisheries and recreational activities. The limited reversibility of eutrophied lakes, which invokes high societal costs, has motivated the development of environmental policies and management measures. These follow precautionary principles limiting nutrient inputs and imposing nutrient load limits, for example, by upgrading wastewater treatment plants, regulating agriculture, forestry, and urban stormwater management, and limiting atmospheric deposition (e.g., (72)). Reversing hysteretic eutrophication requires combinations of measures, including biomanipulation to increase algae grazing, binding or removing phosphorus-loaded sediments, and water level manipulation (72, 73).

### 2.2.2 Forest–savannah–grassland transitions: multiple drivers, spatial dynamics, and bistability

#### The phenomenon and its characteristics

In the tropics and subtropics, forests, savannahs, and grasslands follow rainfall gradients (74, 75). At intermediate rainfall, these landcover types can occur under the same climate, evidence of alternative stable states maintained by fire-vegetation feedback loops: In the absence of tree shading, grasses grow quickly and fuel fires that kill tree seedlings, while closed canopies shade out grasses and suppress fire. The transitions from one state to another are driven by several factors, including precipitation levels, wildfire incidence, soil conditions and deforestation (75, 76). In temperate prairies, fire suppression similarly lets shrubs take over grassland (77). In large rainforests such as the Amazon, evapotranspiration recycles moisture and maintains rainfall, a further stabilizing feedback (78, 79). Warming, drought, land-use change, infrastructure, mining, and deforestation weaken these feedbacks and strengthen forest-declining ones (79, 80).

These transitions are broadly discontinuous (often labelled “catastrophic”), exhibiting (albeit disputed) bistability (77). Furthermore, the formation of spatial patterns is an important factor that can mediate tipping dynamics and even evade them altogether (81). The spatial extent of these transitions ranges from localised ecosystems to continental biomes, such as the prairies of the Central Great Plains, USA (77), the Amazonian rainforest (74, 75, 79), or the Congolian rainforest (74, 75). Much research has focused on the Amazon, where there is limited evidence for a single, system-wide tipping point (82). However, the scale and persistence of pressures can push the system beyond safe limits even without sharp thresholds, and local shifts can cascade through weakened moisture recycling (79, 83, 84). These transitions are hard to reverse on human timescales, and exhibit strong hysteresis (79), because when specifically adapted tree species disappear, the emergence of new tree species would take hundreds to thousands of years.

### Models

These transitions are often analysed as standard discontinuous saddle-node bifurcations (77, 85), as explained in section 2.1.4. More recent work shows a strong dependence on the exact set of environmental and biotic factors, and models describe these phenomena as heteroclinic oscillations (i.e. transient, sequential switching) between stable biomes, Hopf bifurcations, or limit cycles, in addition to the standard saddle-node bifurcation (76, 86). Thus, instead of depending on changes in a single parameter, most systems respond to changes in multiple, inter-dependent parameters. Therefore, tipping points are better described as 'tipping surfaces' in a multi-dimensional parameter space (11, 87). The complexity of large rainforest systems, in particular in relation to their role in the Earth system, are also captured by Earth system models and dynamic global vegetation models that represent biophysical processes between vegetation and atmosphere, rather than using relatively simple dynamical equations (88).

### Attempts at empirical evidence, risks, and management options

Evidence of forest-savannah transitions and of forest encroachment on savannahs has been observed across several tropical areas (89, 90). For entire, large rainforests, such as the Amazon, statistical early warning signals exist (91), while observational evidence for tipping points is limited to localized regions (92). Modelling studies, paleoclimatic evidence, and expert assessments agree that critical values of global warming and deforestation are approaching fast, and the combination of global warming and deforestation means that each individual threshold is further reduced when both pressures act at the same time (84, 88).

Besides halting climate change by cutting greenhouse gas emissions, ending deforestation and investing in reforestation programs can restore the weakened atmospheric moisture recycling feedback. Once deforested, forests lose their ecosystem services for decades. However, they can be restored and biodiversity can, depending on the location, revitalize the ecosystem sooner than previously assessed (93). Otherwise, continued loss of major rainforests including in the Amazon, Congo, or Indonesia, erode biodiversity, regional rainfall, and the dependent agriculture, with effects potentially reaching global supply chains (94).

### 2.2.3. Earth-system tipping elements: interactions, cascades, and cross-scale dynamics

#### The phenomenon

Tipping elements in the Earth system are important subsystems of the Earth where at least one decisive feedback can drive the system to an alternative state beyond its tipping point (3, 55). For instance, fire and tree growth feedbacks are decisive for state changes in the Amazon rainforest (95), the salt-advection feedback for the AMOC (Atlantic Meridional Overturning Circulation (96) or the melt-elevation feedback for the Greenland Ice Sheet (97). Like ecological regime shifts, which can tip before reaching their own climatic thresholds (98), tipping elements in the Earth system interact across scales in space and time (99). For instance, meltwater from the Greenland Ice Sheet enters the North Atlantic and can slow down the AMOC that itself shifts atmospheric rainfall patterns such that the southern Amazon rainforest may receive more rainfall, while the northern rainforest region may receive less. The AMOC is a particularly important tipping element with ample evidence that it can impact both polar ice sheets on Greenland and Antarctica, as well as atmospheric patterns (e.g. El Niño Southern Oscillation).

#### Models

Early models of interacting tipping elements originated from the studies of shallow lakes (100). They modelled tipping elements as simple bistable systems with saddle-node and/or Hopf bifurcations dependent on the type of tipping element and interactions as simple parameterized functions (e.g., (101, 102)). Later work used complex network approaches for larger sets of interacting tipping elements but oftentimes lacked the underlying physical interaction processes that were instead assumed linear (103). More recent models include both simple but physically grounded representations of dynamical systems equations for tipping elements and their interactions, also considering the possibility for nonlinear interactions between tipping elements (e.g. (104, 105)). Building on early advancements (106), studying the dynamics of tipping elements in complex Earth system- and domain specific models is a rapidly developing field (e.g. (107–109)). More detailed Earth system models study interactions directly (e.g., (110, 111)), and their output can inform simpler models, combining mechanistic clarity (98) with realism.

### 2.3 Social and socio-technical transitions: agency, social influence, and institutions

The presented social examples illustrate how transitions theory affects interpretation and measurements in social transitions. Neighbourhood segregation is a foundational case of social tipping, which aims to explain how individual behavior drives transitions in collective patterns. The segregation model has traveled from the social sciences into physics, where it is now studied as a phase-separation problem (112). Social norm change and technology adoption show how social influence can change individual behaviour and generate non-linear collective outcomes. Finally, social-ecological traps describe how persistent poverty and environmental degradation form and persist through mutually reinforcing social and ecological feedbacks. These approaches shift how we understand poverty and development to an emergent outcome of system-level interactions.

### 2.3.1 Neighbourhood segregation: individual behavior and emergent collective patterns

#### The phenomenon

Segregation served as one of the early examples of how micro-level behavior and preferences of individuals could lead to unintuitive – and often undesirable – macro-level phenomena in populations. The classic case is 'white flight' from US neighbourhoods in the 1940-1970s, when, amid government policies that enforced segregation (113), white residents left once the share of black residents passed a perceived threshold, coined as tipping point (114). To this day, most major U.S. metropolitan areas are highly segregated, which is associated with inequities in income, education, employment, access to infrastructure and services, and political polarization (115). Models show that segregation can emerge even when everyone prefers integrated neighborhoods (116), making it particularly hard to reverse.

#### The model

Segregation models, first developed by Sakoda and later by Schelling, are game-theoretical checkerboard models. They show how segregation can result from even weak preferences for like neighbors (117–120). For example, even when agents have no preference for having like neighbors or prefer having both like and different neighbors, preferring to not be in the minority is enough to generate non-linear segregation dynamics, with segregation increasing disproportionately with preferences for like neighbors. However, Schelling notes that predicting this tipping point, or the threshold percentage of minorities at which a neighborhood will flip, is not trivial based on the average or modal individual preferences. Whether and at which threshold tipping occurs depends on the distribution of other-group tolerances, initial population ratios, and the movement rules of individuals (121). However, despite the uncertain threshold, neighborhood tipping is typically described in the modeling and empirical literature as discontinuous and irreversible (114, 122); that is, once white residents have evacuated a neighborhood, the neighborhood rarely changes back.

### Attempts at determining the threshold

Early attempts to determine whether a neighborhood tipping point exists concluded there was no universal or predictable tipping point (e.g., (123)), though later studies using regression discontinuity methods showed that city-specific tipping points exist yet are associated with racial attitudes that have also changed over time (122). More recent data suggests that between 1970 and 2010, tipping became less associated with racial attitudes and more strongly with access to housing options outside of the neighborhood, suggesting mechanisms other than the ones described in Schelling's model (121).

### Path forward

There have been many critiques of Schelling's focus on the relationship between individual preferences and racial segregation, while not accounting for asymmetries in wealth, social and economic power differences between racial groups (124), racist policies and structural and systemic biases in legal, financial, educational and other systems that exist in the U.S. (113). This includes housing ordinances that limited minority entrances to predominantly white neighborhoods to prevent exceeding the critical threshold and triggering white flight (123). Therefore, while the model illustrates that unexpected collective dynamics can emerge from very subtle changes in individual preferences, translating model insights to empirical contexts requires careful analysis and interpretation of the context and variety of factors that shape human behavior.

### 2.3.2 Social norm change and technology adoption: social influence and collective change

#### The phenomenon

There is evidence across several domains, from health behaviors to technology adoption, that people look to others when making decisions for themselves (125). Social interdependence can arise for many reasons, ranging from imitation to social learning, reputational considerations, group identification, conformity to social norms, or benefits that come with coordination. It is also self-reinforcing (126): as more people adopt a belief or behaviour, it becomes more socially rewarded and desirable. As such, social interdependence can reinforce the status quo, but it can also be a force for rapid social change (126, 127), potentially leading to tipping points. Lab experiments on the emergence of social conventions find that a committed group can overturn an established convention (128), though the necessary size of the committed minority depends on features such as network structure (33, 129, 130).

Social dynamics and peer effects have been used to explain patterns of change in real world behavior such as adoption of distributed low-carbon energy technologies (131, 132). Early adopters of technologies, such as rooftop solar panels or electric vehicles, serve as visible social proof that a technology is viable, lowering the perceived risk for subsequent adopters (131). Empirical data on California households indicates that a single additional installation raises the probability of a subsequent adoption by close to a percentage point, a peer effect attributed to greater awareness, reduced perceived risk, and social visibility (132). A similar pattern of geographically decaying imitation appears among German households (133), suggesting the effect is not specific to any single market, culture, or policy regime.

### Threshold Models of Collective Behavior

A popular early conceptual framework, rooted in rural sociology, describes adoption times as bell-shaped, so that cumulative adoption follows an S-curve; adopter categories, from innovators to laggards, differ in traits such as risk tolerance (134). Bass (135) later formalized this in management science with a model equivalent to the Susceptible-Infected model, where contagion happens through peer influence. As technologies vary in their upfront risks and their susceptibility to peer influence, the costs and benefits of adopting depend, in heterogeneous ways, on the number of others who have already done so. Granovetter's threshold model of collective behavior provides a formal micro-foundation to explain non-linear patterns of aggregate technology adoption or social norm change (136). Granovetter proposed that individuals, faced with a binary choice, make this decision based on an idiosyncratic threshold that determines when personal benefits of engaging in a behavior outweigh the costs, and depends critically on the number of others who have already adopted the behavior. The model shows that small changes in the distribution of individual thresholds can produce discontinuous collective outcomes, even with no changes in average preferences. When thresholds are distributed normally, the cumulative adoption curve follows Rogers' empirically determined S-Curve. However, small changes to the distribution of thresholds can produce dramatically different collective outcomes, ranging from full adoption to no adoption (130). A system can remain in an apparently stable low-adoption state until a critical mass of adopters leads it to cascade rapidly toward saturation. The checkerboard segregation models described above are a spatially explicit version of the same logic.

### Policy Implications

One implication of considering social interdependence in technology adoption is that policymakers can leverage these social dynamics to bring about tipping points in technology adoption by strategically seeding new behaviors (4, 127, 137, 138). This "bottom-up" approach might be especially important during periods of limited action among political elite. However, growing polarization around climate change, including low carbon behaviors, also means that interventions, if not carefully conceived, could lead to backlash or counter-movements (129, 139). Finally, the literature on social tipping points has been criticized for using the term liberally (9), and some uses of tipping point terminology can lead to unethical outcomes, such as reinforcing colonial and extractivist practices that pushed the system across the tipping point in the first place (140).

### 2.3.3 Social-ecological poverty traps: coupled feedbacks, persistence, and institutions

#### The phenomenon

Poverty traps are stable equilibria, maintained by self-reinforcing cycles of poverty at individual, community or country level (141–143). The persistence of poverty traps implies a level of hysteresis that makes escaping from this undesirable state much harder than it is to "fall" into it due to gradual wealth loss or small perturbations. The study of poverty traps focuses on the mechanisms that lead to poverty and that prevent a transition out of such states. Social-ecological poverty traps focus on the interactions between the natural environment and poverty and assume that these traps occur when environmental decline and poverty reinforce each other (144–146). Here, emphasis is on the role of social and environmental shocks, the importance of adaptiveness, natural capital, and diversity in withstanding shocks, cross-scale drivers of poverty such as political regime shifts or global market fluctuations, and the role of mismatches between short-term economic gains and long-term ecological health in driving the emergence of traps (145, 147).

Different characteristic feedback mechanisms have been classified into trap archetypes based on potential (accumulated assets and resources), connectedness (strength and density of interactions), and resilience (in the ecological sense, the capacity to absorb disturbances and reorganize while retaining the same identity) (144, 145, 147). Here, poverty traps emerge from low resilience, low connectivity, and low potential in the system (144, 147). In contrast, rigidity traps are characterized by high resilience, potential, and connectedness: they resist change even when it is needed, which erodes adaptive capacity and raises the risk of collapse (147, 148). Examples of rigidity traps include modern industrial agriculture with intensive use of technology and chemicals that degrade the underlying natural capital (like soil health), making the system highly vulnerable to shocks (146), or highly specialized communities that produce or extract a single high-value resource at the expense of diversity, such as the Maine lobster fishery, a 'gilded trap' whose monoculture is threatened by rising ocean temperatures (149–151).

### Models

Bifurcation theory was used in economics to explain poverty traps starting in the 1950s, which was often interpreted to imply the need for large asset inputs to "push" households or economies out of the trap (141, 152), although empirical evidence provides more nuanced interpretations (see below).

Other theories of poverty focus primarily on institutions, drawing on evolutionary game theory to understand why unequal and poverty-producing institutions persist. In these works, repeatedly interacting agents choose between strategies to attempt maximizing their utility, which is shaped by incentive structures and the dynamic choices of other agents (126). These models show that poverty can persist because it is costly to 'cooperate' (i.e., contribute to a social good) within an environment of mostly 'defecting' individuals or organizations, who free ride others' contribution to the social good (153), or to coordinate to reform unequal institutions (154).

On the individual and community level, models of neighborhood effects, drawing on checkerboard models like those above, conceptualize poverty traps as a coordination problem arising from poverty-enforcing social norms that create little social reward for actions that could lead out of poverty (155). Finally, social-ecological traps models, often using coupled dynamical systems models, conceptualize poverty traps as attractors in the stability landscape. They offer a unified approach to understanding mechanisms of poverty traps at different levels, while also accounting for feedbacks between the environment and human systems that can generate otherwise unanticipated consequences for development (156–158). Changing feedback mechanisms can give rise to fundamental transformations that reconfigure system structures and alter the stability landscape (146), which is the focus of social-ecological and sustainability transformations discussed below.

#### Policy implications

Addressing persistent poverty requires understanding and undoing the reinforcing feedback mechanisms through interventions that pull systems out of this state, for example by addressing the drivers of poverty at national, community, and household levels (158). Important work studying poor communities with evidence from randomized control trials has shown that specific interventions, such as access to healthcare, education, information, savings and financial services, as well as addressing uncertainty, are needed to lift people out of poverty (159). Furthermore, addressing the linkages between rural poverty and infectious diseases has been shown as a key feedback cycle in a modeling study (157).

### 2.4 From socio-technical transitions to sustainability transformations

Large-scale, systemic societal transitions are much broader, multi-scale, and multi-system, where innovations emerging at the local level that can be progressively adopted into new regimes. When transition processes are driven by intentional interventions, normative dimensions come into focus, and the question of who decides what to transition to, and how. While more formal transitions theories can explain mechanisms and dynamics, its original focus on single-variable outcomes leaves space for developments encompassing the multiple actors, interests, morals and power dynamics involved in societal and social-ecological transitions.

### 2.4.1 Energy transitions: niches, regimes, and socio-technical pathways

How to decarbonize our energy systems to mitigate climate change is a central focus of the environmental social sciences. This requires large-scale, fundamental changes in infrastructure and technologies, markets, spatial planning, social and political systems, consumer behavior, social norms, and cultural identities that underpin our societies. This makes the energy transition an inherently interdisciplinary research field. Taking a critical transitions lens can help to understand why the current system doesn't change easily, and what conditions might trigger a rapid transition.

#### The model: The Multi-Level Perspective

The multi-level perspective (MLP) is an influential approach from the socio-technical transitions literature that tries to capture systemic change processes. It focuses on how multi-level processes interact to enable or inhibit large-scale technological systems change (160, 161). It conceptualizes how existing regimes, consisting of interactions between technologies, policies, infrastructures, and consumption habits, are locked in, and how innovations emerge in niches. For example, while acting at the periphery of the existing regime that favours, and is influenced by, incumbent actors, such as the fossil fuel industry, community energy systems seek to fundamentally change energy production and use, enabled by decentralized renewable energy technology (162). Niches are the main site of learning and experimentation. Regimes and niches are nested within a broader, slow-changing socio-technical landscape, here including the geo-politics of fossil fuels and oil price fluctuations.

MLP mostly draws on transition theory through metaphor, relying on qualitative analysis of case studies rather than formal modeling that could anticipate when a new regime will overtake an existing one. However, there are many parallels between MLP and formal transitions theory, offering a potential move from metaphor to analogy by taking inspiration from the physical models above or those describing adaptive processes in ecosystems (163). For example, socio-technical transitions take place through micro-level changes ("niches"), in which interaction effects may play a role to generate macro-level system states ("regimes") controlled by multiple parameters (technologies, policies, infrastructures, habits, and discourses) that are susceptible to external fields ("landscape") and perturbations (external shocks, such as oil price fluctuations). Lock-ins of infrastructural sunk costs, habits and behaviours, and entrenched political systems give rise to stabilizing feedbacks that maintain incumbent regimes and lead to hysteretic effects.

### Other models

A strand of transition management embraces societal systems as complex adaptive systems, with coevolution, self-organization, and attractors (164, 165), describes transitions as S-shaped sequences of predevelopment, take-off, acceleration, and stabilization (166), and identifies recurring patterns along transition paths (167). A smaller modelling strand also exists (168), including threshold models (169) and agent-based models that reproduce historical pathways such as the shift from sail- to steamboats (170).

### Attempts at empirical evidence and policy implications

Empirically, MLP has been used to understand why different energy transition pathways are observed in different contexts (161, 171). For example, despite similar starting points, offshore wind boomed in the UK in the early 2000's but stagnated in the Netherlands. The UK's success is attributed to a proactive 'system builder', the Crown Estate, within powerful public-private networks, while the Netherlands tried to push innovations within existing rules (172). The MLP has also shaped policy, informing the transition-management approach that entered Dutch energy policy and steering it toward supporting niche innovations, although the programme was found at risk of capture by incumbent interests (173). The framework has been criticized for not adequately addressing the power dynamics of existing regimes, though more recent efforts aimed to better account for how the incumbent regime actively inhibits transition pathways (174) and theorize on the role of power in sustainability transitions (165, 175–177).

### 2.4.2 Social-ecological and sustainability transformations: agency, plurality, justice, and alternative futures

#### The phenomenon and its characteristics

In the context of biodiversity loss, environmental degradation, climate change, and social inequality and injustice, social-ecological and sustainability transformations (short: 'transformations') research starts with a recognition of the unsustainability of the current system's state and centers the exploration of what could be instead (178–180). With roots in sustainability science and social-ecological resilience research, transformations are conceptualized as a fundamental restructuring of the set of relationships characterizing a particular state or development pathway (181, 182), and how they connect to ecological systems across scales (178). Transformation involves ethical interventions aimed to achieve more equitable outcomes (183) that rely on the role of key actors, whose interactions and distributed agency can help trigger change or stabilize new regimes (184). The attractor - or desirable state – emerges through an exploration of imagined futures, including its material flows, social structures, narratives, and practices that define different regimes, interrogating what a new regime should look like and who benefits and who loses (178).

### The model/theoretical framework

Similar to socio-technical transitions described by the MLP, transformation processes are conceptualized to involve changes in the practical sphere (e.g., behaviors, technologies, management), the political sphere (social systems, rules, power structures), and the personal sphere (beliefs, values, and worldviews) (185). These spheres are interlinked, together forming complex adaptive systems, where both, human behaviour and imagined, possible futures, co-evolve with their socio-cultural and biophysical context (178, 186). Distinct from socio-technical transitions, the desirable state that the system is to transition to is not defined and is not associated with a specific (technological) transition pathway. Rather, transformations research focuses on the pluralistic and contested nature of sustainability instead of assuming that there is a single desirable goal (187). It moves away from the dystopian views of impending catastrophe towards building on the current contexts and historical pathways (188, 189). Here, agency plays a key role in terms of who imagines, who controls, and who initiates collective imagination processes (178).

Developing transformation pathways requires higher-level learning that moves beyond problem-solving and incremental improvements to existing strategies, rules and parameters, and includes questioning assumptions and strategies, and assessing core beliefs, worldviews, and values (190). Leverage points are places in a system where a small intervention can trigger system-wide change (191). Recognizing the complexity, path dependence, and stochastic nature of systemic change, transformation research emphasizes the need for enabling conditions (192), existing initiatives that can serve as seeds in fostering a desirable future (189), and embedded forms of governance that can help navigate transformation processes and ensure that desirable processes and outcomes are strengthened vis à vis undesirable ones (193). The implementation of transformation pathways involves a 'preparation' phase of weakening existing structures and self-reinforcing feedbacks that maintain the dominant regime, a navigation phase between the old and new regime, and a stabilization phase of establishing new structures and feedbacks that build resilience (182), similar to the phases of the adaptive cycle (147). The rezoning of Australia's Great Barrier Reef Marine Park illustrates these phases: a flexible agency initiated the transition, mobilized scientists and the public, and secured political backing at critical moments (194) .

### Path forward

Transformations scholars often use a participatory action approach to research, that is, they aim to solve on the ground problems in collaboration with communities and non-academic partners. Such co-design is rarer in formal transitions research. Strengthening the conversation between mechanistic understanding of formal transitions with participatory transformations research that inquires about meanings, values, agency, and context, can push the frontiers in both research fields (12), as we further discuss below.

## 3. Learning across scientific fields

The transition phenomena presented above share several features, but no single set of properties defines them all (Fig. 1). Thresholds appear in many cases, which, once crossed, lead the system into alternative system states, but their empirical detectability and usefulness as management targets remain debated (e.g., (7, 195)). Likewise, abruptness, criticality, hysteresis, irreversibility, and predictability do not typically occur together: a transition may be continuous or discontinuous; a threshold may be clearly defined in a model but blurred by stochasticity in an empirical system.

In ecological and social-ecological systems, attention also shifts toward the feedbacks that stabilize regimes, spatial and temporal dynamics, cross-scale interactions, and the possibility of anticipating transitions through early-warning signals. More recent work has expanded the focus further toward risks and consequences, agency and power, the normative implications of deliberately triggering 'positive' transitions, and the role of imagining alternative futures.

Across the cases, we see a broad expansion in what research on transitions seeks to explain. Many mechanisms and mathematical ideas developed in one domain remain useful in others. New features become important as the systems under study become more complex and more strongly shaped by environmental and social context. As our discussion above shows, some transitions features are well captured by theoretical models while others feed ongoing debate and stimulate new research.

Classical examples in physics, such as ferromagnetism and phase changes of water show how interactions among many microscopic components can produce qualitative changes in aggregate system behaviour. They introduce fundamental concepts, such as order and control parameters, criticality, bifurcations, susceptibility, universality, and early-warning behaviour. Models of collective failure add further elements, including heterogeneity among components, cascading responses, instability, stochasticity, and irreversibility. Here, assumptions are clearly specified, the relevant interactions can often be isolated, and associated thresholds can be tested experimentally in controlled laboratory environments. This makes it possible to derive general results.

The extension from physical systems to epidemics shows both the strength and the limits of this formal approach. Percolation theory and network models show how large-scale collective behaviour can emerge from the pattern of connections among individual units, but network structure can strongly affect transition behaviour, including whether a clear epidemic threshold exists. This provides an important lesson for transitions research more broadly: similar mathematical structures can appear in very different systems, but results from one model cannot simply be transferred to another system without checking whether the underlying assumptions hold.

In Earth- and ecological systems, the challenge is often not only to understand how a specified set of interacting components behaves, but also to determine which processes, drivers, and system boundaries are most important. Lake eutrophication, for example, depends on reinforcing ecological feedbacks, but also on lake depth, circulation, spatial heterogeneity, nutrient dynamics, and environmental fluctuations. Whether critical transitions in lakes manifest also depends on the timescale to define equilibrium. Forest–savannah–grassland transitions involve interactions among vegetation, fire, precipitation, land use, and spatial patterning. At larger scales, Earth-system tipping elements research examines how tipping in one element can affect the behaviour and stability of others, akin to the smaller-scale ecological studies on trophic cascades.

As a result, a single control parameter or a single threshold may no longer be enough to describe the system. Multiple interacting drivers may instead form tipping surfaces or critical manifolds, and transitions may spread or be modified across space and scales. The advantage is a richer description of the mechanisms operating in real environments. The difficulty is weaker experimental control, greater reliance on observations and simulations, and more uncertainty about causal mechanisms and thresholds. These ecological and Earth system examples illustrate how difficult it can be to match an empirical system to a specific theoretical class of transition.

Social and socio-technical transitions introduce another important feature: the interacting elements are people and organizations whose behaviour responds to incentives, social influence, identities, institutions, expectations, and the behaviour of others. Formal transition mechanisms remain useful. Schelling-type segregation models and Granovetter-type threshold models, show how individual decisions can generate nonlinear and sometimes unexpected collective outcomes.

Social norm change and technology adoption demonstrate how social influence can reinforce behaviours or support rapid collective change.

However, once the interacting elements are agents, differences among them become important. Individuals and organizations differ in their goals, resources, opportunities, and ability to influence others, which can shape both whether a transition occurs and who benefits from it. Neighbourhood segregation, for example, cannot be understood only through individual preferences; institutional structures, discriminatory policies, unequal wealth, and differences in social and economic power also shape the observed patterns. Socio-technical change similarly depends not only on individual adoption, but also on technologies, infrastructures, markets, institutions, incumbent actors, and new entrants. The focus therefore expands from explaining collective patterns to understanding how agency, institutions, and power shape both the process and the outcomes of transitions.

In the final expansion on sustainability transformations, explaining change processes is not enough, but the direction and desirability of change also become central questions. Socio-technical transition research broadens the focus from individual behaviour to niches, regimes, and landscapes, and associated infrastructures, policies, and pathways of system change. Sustainability transformations go further by recognizing that there may be several alternative futures and that different groups may value these futures differently. A transition that benefits one group, sector, or scale may create risks and costs for another. Therefore, agency, power, justice, and the ability to imagine and shape alternative futures become part of the analysis. The object of study is no longer only movement between (pre)defined system states, but also the reorganization of relationships, institutions, practices, material flows, and power structures that enable and maintain those states. Researchers and other knowledge producers may also take part in transformation processes rather than only observe them from outside.

Taken together, these four domains (Figure 1) show a recurring trade-off in transitions research. More abstract formal models can identify mechanisms and patterns that appear across very different systems, but they do so by leaving out features that may be important in specific cases. As theories include environmental context, scale, agency, institutions, power, and alternative futures, they can better represent real-world transitions. However, they become harder to generalize, test experimentally, and predict precisely. The main challenge is therefore not to find one best theory of transitions, but to understand where particular approaches are useful, where their assumptions fail, and how ideas can move across fields without losing their meaning.

This also points to where cross-field learning may be most useful. Formal transition theory can help social, socio-technical, and transformation research develop clearer hypotheses about feedbacks, criticality, path dependence, cascading change, and nonlinear responses. In the other direction, social and sustainability research can broaden formal approaches by asking whose outcomes are represented, how agency and power shape system dynamics, and whether apparently 'positive' transitions are beneficial across different groups, places, and scales. The value of integration lies less in building a universal theory than in making clear how phenomena, assumptions, mechanisms, and methods relate to each other. Such similarities may extend beyond metaphor: material fracture, ecosystem reorganization, and social polarization can share mathematical structure, so identifying where analogies inform and where they break down is a key cross-disciplinary task. The co-citation analysis below examines how far this exchange across fields has already occurred and which areas of research have helped connect these scientific communities.

## 3.1 Co-citation network analysis

The differences and similarities identified above are a result of how terms, concepts, and theories have migrated across different research fields. The co-citation network analysis shown in Figure 2 is based on literature extracted for the topic modeling discussed in the introduction, focusing on the five transitions terms excluding phase transitions. It illustrates how transitions knowledge domains are linked more broadly, assessed above through selected examples, and how different (sub-) disciplines underly the patterns of interdisciplinarity over time.

Visualized for the most highly cited articles in Fig. 2, our analysis shows that before 2005, the network was mostly comprised of the EES domain centered on highly cited articles treating the pacific climate regime shift (196–198), with a small, separate cluster comprised of less than a handful of articles centered on socio-technical transitions (most highly cited article: (160)). A decade later, the share of cross-disciplinary edges had nearly doubled. The socio-technical transitions community, while highly cited within its own cluster, only slightly grew in number of articles, and became connected to the EES domain through one significant bridge (148), while the remaining nodes formed a densely connected cluster spanning several domains, including EES, geography, and mathematics. In this period, cross-disciplinary co-citations were driven by literature on early warning signals (199, 200). This network pattern remains through 2025, with several new, small clusters appearing on the periphery of the network, and a new bridge emerging (201) linking both key clusters and marking the visible entry of the sustainability transformations field. Other articles function both as highly influential (high citation counts) and as bridges across multiple communities (high betweenness centrality) (55, 161, 181, 199, 202).

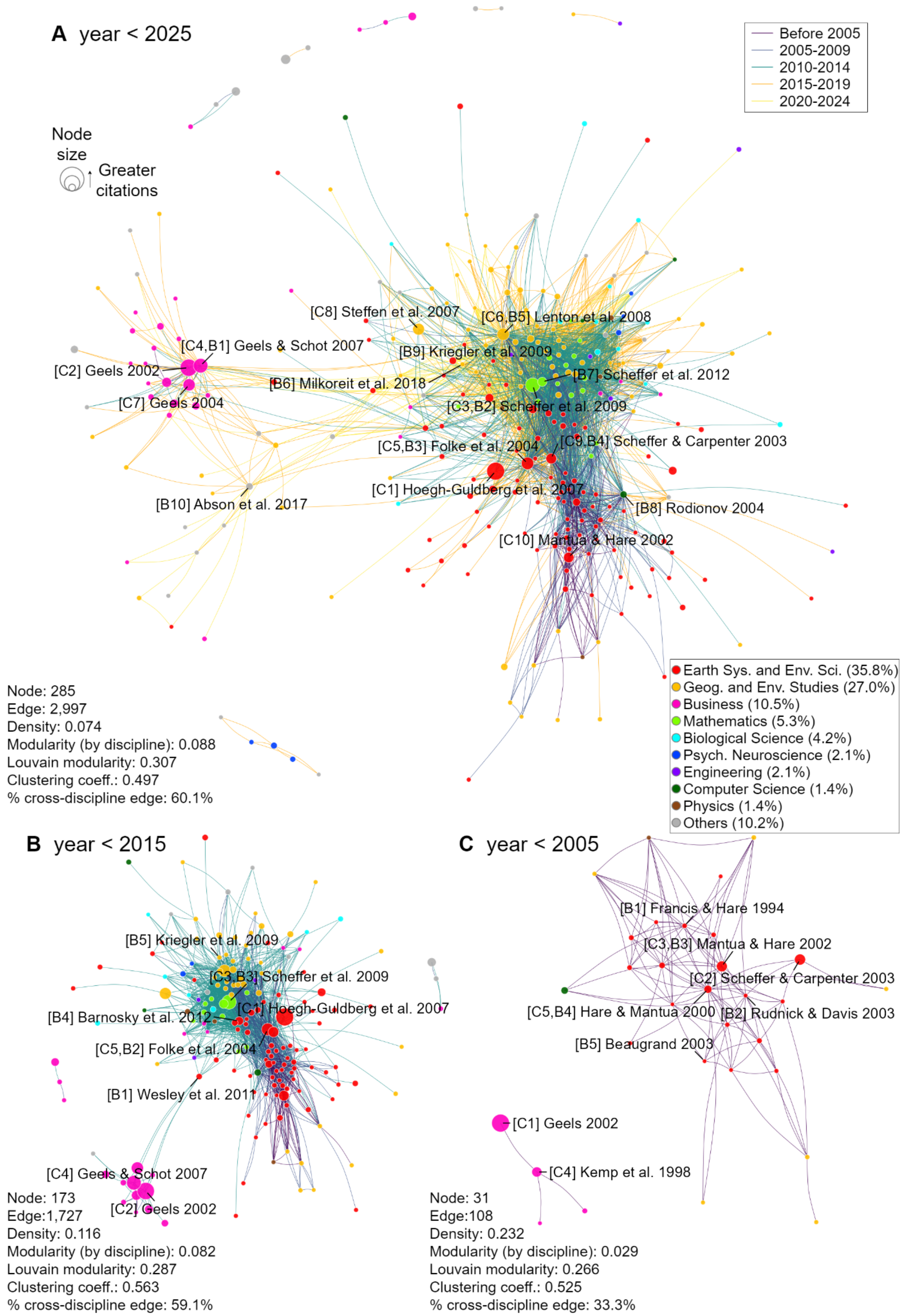
A year < 2025
Before 2005
2005-2009
2010-2014
2015-2019
2020-2024
Node size
Greater citations
[C8] Steffen et al. 2007
[C6,B5] Lenton et al. 2008
[C4,B1] Geels & Schot 2007
[B9] Kriegler et al. 2009
[C2] Geels 2002
[B7] Scheffer et al. 2012
[B6] Milkoreit et al. 2018
[C7] Geels 2004
[C3,B2] Scheffer et al. 2009
[C9,B4] Scheffer & Carpenter 2003
[C5,B3] Folke et al. 2004
[C1] Hoegh-Guldberg et al. 2007
[B10] Abson et al. 2017
[B8] Rodionov 2004
[C10] Mantua & Hare 2002
Node: 285
Edge: 2,997
Density: 0.074
Modularity (by discipline): 0.088
Louvain modularity: 0.307
Clustering coeff.: 0.497
% cross-discipline edge: 60.1%
Earth Sys. and Env. Sci. (35.8%)
Geog. and Env. Studies (27.0%)
Business (10.5%)
Mathematics (5.3%)
Biological Science (4.2%)
Psych. Neuroscience (2.1%)
Engineering (2.1%)
Computer Science (1.4%)
Physics (1.4%)
Others (10.2%)
B year < 2015
[B5] Kriegler et al. 2009
[C3,B3] Scheffer et al. 2009
[B4] Barnosky et al. 2012
[C1] Hoegh-Guldberg et al. 2007
[C5,B2] Folke et al. 2004
[B1] Wesley et al. 2011
[C4] Geels & Schot 2007
[C2] Geels 2002
Node: 173
Edge:1,727
Density: 0.116
Modularity (by discipline): 0.082
Louvain modularity: 0.287
Clustering coeff.: 0.563
% cross-discipline edge: 59.1%
C year < 2005
[B1] Francis & Hare 1994
[C3,B3] Mantua & Hare 2002
[C2] Scheffer & Carpenter 2003
[C5,B4] Hare & Mantua 2000
[B2] Rudnick & Davis 2003
[B5] Beaugrand 2003
[C1] Geels 2002
[C4] Kemp et al. 1998
Node: 31
Edge:108
Density: 0.232
Modularity (by discipline): 0.029
Louvain modularity: 0.266
Clustering coeff.: 0.525
% cross-discipline edge: 33.3%

**Figure 2:** Co-citation patterns of critical transitions research across disciplinary boundaries. **A | Co-citation network up to 2024** filtered by high citation counts (top 5%) and at least five co-citing articles. Each node is a co-cited article (N=285) weighted by citation counts, and edge is co-cited connections (2,997 edges). Each edge records the publication year of the earliest paper that cited both articles. Node colors represent research field. Labeled articles rank in the top 10 by citation count [C1–C10] or betweenness centrality [B1–B10]. **B | Filtered co-citation network up to 2014** from the panel A network. **C | Filtered co-citation network up to 2004** from the panel A network (N=31 articles). In panels B and C, the five most cited articles and those with top-5 highest betweenness centrality are labeled. Detailed analysis methods, network metrics, and the full lists of references are provided in the SI.

## 4. Summary Points and Future Issues

A wide range of theories explains how systems reorganize, persist, tip, or transform. Across the examples reviewed here, these theories describe different mechanisms, from interactions and feedbacks to network structure, cross-scale dynamics, human agency, institutions, and the imagination of alternative futures that can serve as crystallization points for new resilience regimes to emerge. A central challenge is not simply to develop new models, but to determine which theoretical framework is appropriate for a given empirical system.

Real-world systems rarely fit neatly into a single class of transition. The same phenomenon may be consistent with several theoretical descriptions, while in other cases existing theory may not capture the relevant dynamics at all. Testing whether an empirical system meets the assumptions of a particular framework can require large amounts of data. This problem is made more difficult by limits to observation and measurement. Complex systems are often monitored at spatial or temporal scales that are too coarse to identify the mechanisms, thresholds, or warning signals of interest. In ecological and Earth systems, long timescales and limited observations often require indirect approaches, while spatial structure can delay, spread, or alter transitions. Improving prediction therefore requires both better observations and theories that account for scale, space, uncertainty, and interactions among systems.

A major research opportunity lies in stronger exchange between formal transition science and research on social-ecological and sustainability transformations. Formal models provide clear ways to study mechanisms, feedbacks, criticality, path dependence, and nonlinear change. Transformation research adds dimensions that formal models often represent only weakly, including agency, institutions, power, justice, multiple outcomes, and different views of desirable futures.

Progress is therefore unlikely to come from a single universal theory of transitions. Instead, it will depend on understanding where different approaches are useful, where their assumptions break down, and how they can complement one another. A productive synthesis would retain the clarity and testability of formal theory while improving our ability to represent coupled social-ecological systems and the different consequences of deliberate change. Such integration can strengthen both our capacity to anticipate harmful transitions and our ability to guide transformations toward more sustainable and just futures.

## Author contributions

All authors reviewed and edited this article. Initial conceptualization: EHK, KH, JCR, GFe, GFa, GIH, SMC; section lead writers – PHYS/MATH: GFe, GFa, AS, JL, JB, GH; - EES: KH, LWE, NW, HJ; - SocSci: NM, SMC; topic analysis: JCR; co-citation network analysis: WO, ERZ; figure creation: XH, WO, EHK; draft writing, reviewing, final editing: EHK & VVV.

## Acknowledgments

This review emerged from a series of international workshops of the Earth Resilience and Sustainability Initiative (ERSI) at Princeton University, Stockholm Resilience Centre at Stockholm University, Potsdam Institute for Climate Impact Research, and the University of Amsterdam. The authors would like to acknowledge the contributions of workshop participants to discussions and responding to a survey distributed in the early stages of the manuscript, including Simon Levin, Alan Hastings, Miguel Centeno, Garry Peterson, Juan Bonachela, Marty Anderies, Nigel Goldenfeld, Maya Almaraz, Florian Ulrich Jehn, André Pinto da Silva, Uno Svedin, Niklas Kitzmann, Wolfram Barfuss, Victoria Junquera, Constantin Arnscheid, Vanessa Weinberger, Mari Kawakatsu, Maria Garcia Otero, and Stefanie Crabtree. We thank Miguel Angel Muñoz for discussions on the initial conceptualization and for valuable input to the final draft. LWE received funding from Formas - a Swedish Research Council for Sustainable Development (2022-02089, 2023-00310, 2023-00321, 2024-01655). JB received funding from the Cooperative AI Foundation. JB and VVV acknowledge support by the Dutch Research Council (NWO) under the grant https://doi.org/10.61686/XAPNR39148 (ROOTS, VI.Vidi.243.009). GIH acknowledges the support of PSC-CUNY TradB Award 56-365, U.S. Department of Commerce National Oceanic and Atmospheric Administration (NOAA Grant NA18OAR4320123) and the National Science Foundation (NSF Grant IOS 2137340). GF received funding from the Lopez-Loreta Foundation and TU Dublin. EHK received funding from an ENLENS grant at the University of Amsterdam.

## Literature cited

1. Earn DJD, Rohani P, Bolker BM, Grenfell BT. 2000. A Simple Model for Complex Dynamical Transitions in Epidemics. *Science (1979).* 287(5453):667–70

2. Hoegh-Guldberg O, Mumby PJ, Hooten AJ, Steneck RS, Greenfield P, et al. 2007. Coral Reefs Under Rapid Climate Change and Ocean Acidification. *Science (1979).* 318(5857):1737–42

3. Armstrong McKay DI, Staal A, Abrams JF, Winkelmann R, Sakschewski B, et al. 2022. Exceeding 1.5°C global warming could trigger multiple climate tipping points. *Science (1979).* 377(6611):

4. Nyborg K, Anderies JM, Dannenberg A, Lindahl T, Schill C, et al. 2016. Social norms as solutions. *Science (1979).* 354(6308):42–43

5. Herrfahrdt-Pähle E, Schlüter M, Olsson P, Folke C, Gelcich S, Pahl-Wostl C. 2020. Sustainability transformations: socio-political shocks as opportunities for governance transitions. *Global Environmental Change*. 63(102097):22

6. Lenton TM, MM, W et al. 2025. Global tipping points report 2025. Exeter, UK

7. Kopp RE, Gilmore EA, Shwom RL, Adams H, Adler C, et al. 2025. 'Tipping points' confuse and can distract from urgent climate action. *Nat. Clim. Chang.* 15(1):29–36

8. Smith SR, Milkoreit M, Geels FW, Lenton TM. 2025. Advancing science, policy and action in tipping points research. *Nat. Clim. Chang.* 15(6):576–77

9. Milkoreit M. 2023. Social tipping points everywhere?—Patterns and risks of overuse. *WIREs Climate Change*. 14(2):

10. Hagstrom GI, Levin SA. 2021. Phase Transitions and the Theory of Early Warning Indicators for Critical Transitions

11. Abrams JF, Pereira LM, Spaiser V, Wunderling N, Villasante S, et al. 2026. Integrating tipping point concepts across diverse systems. *Communications Sustainability*. 1(1):68

12. Rocha JC, Schill C, Lindkvist EAL, Biggs R, Blenckner T, et al. 2026. Regime Shifts and Transformations in Social-Ecological Systems: Advancing Critical Frontiers for Safe and Just Futures. *Annu. Rev. Environ. Resour.*

13. Milkoreit M, Boyd E, Constantino SM, Hausner VH, Hessen DO, et al. 2024. Governance for Earth system tipping points – A research agenda. *Earth System Governance*. 21:100216

14. Cardy J. 1996. *Scaling and Renormalization in Statistical Physics*. Cambridge University Press

15. Binney JJ, Dowrick NJ, Fisher AJ, Newman MEJ. 1992. *The Theory of Critical Phenomena: An Introduction to the Renormalization Group*. Oxford University Press

16. Ewing JA. 1883. III. On effects of retentiveness in the magnetisation of iron and steel. *Proceedings of the Royal Society of London*. 34.220-223:39–45

17. Andrews T. 1869. The Bakerian Lecture.--On the Continuity of the Gaseous and Liquid States of Matter. *Philos. Trans. R. Soc. Lond.* 159:575–90

18. Gross B, Volotsenko I, Sallem Y, Yadid N, Bonamassa I, et al. 2025. The random cascading origin of abrupt transitions in interdependent systems. *Nat. Commun.* 16(1):5869

19. Hemmer PC, Hansen A. 1992. The Distribution of Simultaneous Fiber Failures in Fiber Bundles. *J. Appl. Mech.* 59(4):909–14

20. Zapperi S, Ray P, Stanley HE, Vespignani A. 1997. First-Order Transition in the Breakdown of Disordered Media. *Phys. Rev. Lett.* 78(8):1408–11

21. Kovács K, Hidalgo RC, Pagonabarraga I, Kun F. 2013. Brittle-to-ductile transition in a fiber bundle with strong heterogeneity. *Phys. Rev. E*. 87(4):042816

22. Abaimov SG. 2015. *Statistical Physics of Non-Thermal Phase Transitions*. Cham: Springer International Publishing

23. Pradhan S, Chakrabarti BK. 2001. Precursors of catastrophe in the Bak-Tang-Wiesenfeld, Manna, and random-fiber-bundle models of failure. *Phys. Rev. E*. 65(1):016113

24. Zapperi S, Nukala PKV V., Šimunović S. 2005. Crack roughness and avalanche precursors in the random fuse model. *Phys. Rev. E*. 71(2):026106

25. Unger C, Klein W. 1984. Nucleation theory near the classical spinodal. *Phys. Rev. B*. 29(5):2698–2708

26. Kaiser J. 1950. *Untersuchungen über das Auftreten von Geräuschen beim Zugversuch*. Technische Hochschule München

27. Bonamassa I, Gross B, Kertész J, Havlin S. 2025. Hybrid universality classes of systemic cascades. *Nat. Commun.* 16(1):1415

28. Rundle JB, Klein W, Turcotte DL, Malamud BD. 2000. Precursory Seismic Activation and Critical-point Phenomena. *Pure Appl. Geophys.* 157(11):2165–82

29. Brock WA, Durlauf SN. 2001. Discrete Choice with Social Interactions. *Rev. Econ. Stud.* 68(2):235–60

30. Mann RP, Faria J, Sumpter DJT, Krause J. 2013. The dynamics of audience applause. *J. R. Soc. Interface*. 10(85):20130466

31. Michard Q, Bouchaud J-P. 2005. Theory of collective opinion shifts: from smooth trends to abrupt swings. *Eur. Phys. J. B*. 47(1):151–59

32. Bouchaud J-P. 2013. Crises and Collective Socio-Economic Phenomena: Simple Models and Challenges. *J. Stat. Phys.* 151(3–4):567–606

33. Mittal D, Constantino SM, Vasconcelos V V. 2024. Anticonformists catalyze societal transitions and facilitate the expression of evolving preferences. *PNAS Nexus*. 3(8):

34. Watts DJ. 2002. A simple model of global cascades on random networks. *Proceedings of the National Academy of Sciences*. 99(9):5766–71

35. Liu Q-X, Doelman A, Rottschäfer V, de Jager M, Herman PMJ, et al. 2013. Phase separation explains a new class of self-organized spatial patterns in ecological systems. *Proceedings of the National Academy of Sciences*. 110(29):11905–10

36. Siteur K, Liu Q-X, Rottschäfer V, van der Heide T, Rietkerk M, et al. 2023. Phase-separation physics underlies new theory for the resilience of patchy ecosystems. *Proceedings of the National Academy of Sciences*. 120(2):

37. Cummings DAT, Irizarry RA, Huang NE, Endy TP, Nisalak A, et al. 2004. Travelling waves in the occurrence of dengue haemorrhagic fever in Thailand. *Nature*. 427(6972):344–47

38. Kermack WO, McKendrick AG. 1927. A contribution to the mathematical theory of epidemics. *Proceedings of the Royal Society of London. Series A, Containing Papers of a Mathematical and Physical Character*. 115(772):700–721

39. Grassberger P. 1983. On the critical behavior of the general epidemic process and dynamical percolation. *Math. Biosci.* 63(2):157–72

40. Newman MEJ. 2002. Spread of epidemic disease on networks. *Phys. Rev. E*. 66(1):016128

41. Pastor-Satorras R, Vespignani A. 2001. Epidemic dynamics and endemic states in complex networks. *Phys. Rev. E*. 63(066117):

42. Vázquez A, Moreno Y. 2003. Resilience to damage of graphs with degree correlations. *Phys. Rev. E*. 67(1):015101

43. Serrano MÁ, Boguñá M. 2006. Percolation and Epidemic Thresholds in Clustered Networks. *Phys. Rev. Lett.* 97(8):088701

44. Liu Q-X, Wang R-H, Jin Z. 2009. Persistence, extinction and spatio-temporal synchronization of SIRS spatial models. *Journal of Statistical Mechanics: Theory and Experiment*. 2009(07):P07007

45. Grenfell BT, Bjørnstad ON, Kappey J. 2001. Travelling waves and spatial hierarchies in measles epidemics. *Nature*. 414(6865):716–23

46. Quandelacy TM, Vincente-González MF, Grillet ME, Colomé-Hidalgo M, Herrera D, et al. 2025. Synchronized dynamics of dengue across the Americas. *Sci. Transl. Med.* 17(812):

47. Watts DJ, Muhamad R, Medina DC, Dodds PS. 2005. Multiscale, resurgent epidemics in a hierarchical metapopulation model. *Proceedings of the National Academy of Sciences*. 102(32):11157–62

48. Sun G-Q, Jusup M, Jin Z, Wang Y, Wang Z. 2016. Pattern transitions in spatial epidemics: Mechanisms and emergent properties. *Phys. Life Rev.* 19:43–73

49. Ashwin P, Wieczorek S, Vitolo R, Cox P. 2012. Tipping points in open systems: bifurcation, noise-induced and rate-dependent examples in the climate system. *Philosophical Transactions of the Royal Society A: Mathematical, Physical and Engineering Sciences*. 370(1962):1166–84

50. Kuehn C. 2011. A mathematical framework for critical transitions: Bifurcations, fast–slow systems and stochastic dynamics. *Physica D*. 240(12):1020–35

51. Poincaré H. 1885. Sur l'équilibre d'une masse fluide animée d'un mouvement de rotation. *Acta Mathematica*. 7(0):259–380

52. Thom R. 1972. *Stabilité structurelle et morphogénèse : essai d'une théorie générale des modèles*. Paris: W.A. Benjamin, Inc.

53. Kuznetsov YA. 2004. *Elements of Applied Bifurcation Theory*, Vol. 112. New York, NY: Springer New York

54. Thompson JMT, Stewart HB, Ueda Y. 1994. Safe, explosive, and dangerous bifurcations in dissipative dynamical systems. *Phys. Rev. E*. 49(2):1019–27

55. Lenton TM, Held H, Kriegler E, Hall JW, Lucht W, et al. 2008. Tipping elements in the Earth's climate system. *Proceedings of the National Academy of Sciences*. 105(6):1786–93

56. Alkhayuon H, Tyson RC, Wieczorek S. 2021. Phase tipping: how cyclic ecosystems respond to contemporary climate. *Proceedings of the Royal Society A: Mathematical, Physical and Engineering Sciences*. 477(2254):

57. Hastings A, Petrovskii S, Lucarini V, Morozov A. 2026. Tipping points in complex ecological systems

58. Falmagne G, Stephenson AB, Levin SA. 2026. Interpretable early warnings using machine learning in an online game-experiment. *Proceedings of the National Academy of Sciences*. 123(1):

59. Scheffer M, Hosper SH, Meijer M-L, Moss B, Jeppesen E. 1993. Alternative equilibria in shallow lakes. *Trends Ecol. Evol.* 8(8):275–79

60. van Nes EH, Scheffer M, van den Berg MS, Coops H. 2002. Dominance of charophytes in eutrophic shallow lakes—when should we expect it to be an alternative stable state? *Aquat. Bot.* 72(3–4):275–96

61. van Nes EH, Scheffer M. 2005. Implications of spatial heterogeneity for catastrophic regime shifts in ecosystems. *Ecology*. 86(7):1797–1807

62. Scheffer M, van Nes EH. 2007. Shallow lakes theory revisited: various alternative regimes driven by climate, nutrients, depth and lake size. *Hydrobiologia*. 584(1):455–66

63. Davidson TA, Sayer CD, Jeppesen E, Søndergaard M, Lauridsen TL, et al. 2023. Bimodality and alternative equilibria do not help explain long-term patterns in shallow lake chlorophyll-a. *Nat. Commun.* 14(1):398

64. van Nes EH, Rip WJ, Scheffer M. 2007. A Theory for Cyclic Shifts between Alternative States in Shallow Lakes. *Ecosystems*. 10(1):17

65. van Nes EH, Scheffer M, van den Berg MS, Coops H. 2003. Charisma: a spatial explicit simulation model of submerged macrophytes. *Ecol. Modell.* 159(2–3):103–16

66. Scheffer M, Szabó S, Gragnani A, van Nes EH, Rinaldi S, et al. 2003. Floating plant dominance as a stable state. *Proceedings of the National Academy of Sciences*. 100(7):4040–45

67. May RM. 1977. Thresholds and breakpoints in ecosystems with a multiplicity of stable states. *Nature*. 269(5628):471–77

68. Holling CS. 1973. Resilience and Stability of Ecological Systems. *Annu. Rev. Ecol. Syst.* 4(1):1–23

69. Carpenter SR, Cole JJ, Pace ML, Batt R, Brock WA, et al. 2011. Early Warnings of Regime Shifts: A Whole-Ecosystem Experiment. *Science (1979).* 332(6033):1079–82

70. Pace ML, Batt RD, Buelo CD, Carpenter SR, Cole JJ, et al. 2017. Reversal of a cyanobacterial bloom in response to early warnings. *Proceedings of the National Academy of Sciences*. 114(2):352–57

71. Gsell AS, Scharfenberger U, Özkundakci D, Walters A, Hansson L-A, et al. 2016. Evaluating early-warning indicators of critical transitions in natural aquatic ecosystems. *Proceedings of the National Academy of Sciences*. 113(50):

72. Carpenter SR, Ludwig D, Brock WA. 1999. Management of Eutrophication for Lakes Subject to Potentially Irreversible Change. *Ecological Applications*. 9(3):751

73. Jeppesen E, Meerhoff M, Jacobsen BA, Hansen RS, Søndergaard M, et al. 2007. Restoration of shallow lakes by nutrient control and biomanipulation—the successful strategy varies with lake size and climate. *Hydrobiologia*. 581(1):269–85

74. Hirota M, Holmgren M, Van Nes EH, Scheffer M. 2011. Global Resilience of Tropical Forest and Savanna to Critical Transitions. *Science (1979).* 334(6053):232–35

75. Staver AC, Archibald S, Levin SA. 2011. The Global Extent and Determinants of Savanna and Forest as Alternative Biome States. *Science (1979).* 334(6053):230–32

76. Touboul JD, Staver AC, Levin SA. 2018. On the complex dynamics of savanna landscapes. *Proc. Natl. Acad. Sci. U. S. A.* 115(31):E1336–45

77. Ratajczak Z, Nippert JB, Briggs JM, Blair JM. 2014. Fire dynamics distinguish grasslands, shrublands and woodlands as alternative attractors in the Central Great Plains of North America. *Journal of Ecology*. 102(6):1374–85

78. Staal A, Flores BM, Aguiar APD, Bosmans JHC, Fetzer I, Tuinenburg OA. 2020. Feedback between drought and deforestation in the Amazon. *Environmental Research Letters*. 15(4):044024

79. Flores BM, Montoya E, Sakschewski B, Nascimento N, Staal A, et al. 2024. Critical transitions in the Amazon forest system. *Nature*. 626(7999):555–64

80. Spracklen DV, Baker JCA, Garcia-Carreras L, Marsham JH. 2018. The Effects of Tropical Vegetation on Rainfall. *Annu. Rev. Environ. Resour.* 43(1):193–218

81. Rietkerk M, Bastiaansen R, Banerjee S, van de Koppel J, Baudena M, Doelman A. 2021. Evasion of tipping in complex systems through spatial pattern formation. *Science (1979).* 374(6564):

82. Brando PM, Barlow J, Macedo MN, Silvério D V., Ferreira JN, et al. 2025. Tipping Points of Amazonian Forests: Beyond Myths and Toward Solutions. *Annu. Rev. Environ. Resour.* 50(1):97–131

83. Bochow N, Boers N. 2023. The South American monsoon approaches a critical transition in response to deforestation. *Sci. Adv.* 9(40):

84. Wunderling N, Sakschewski B, Rockström J, Flores BM, Hirota M, Staal A. 2026. Deforestation-induced drying lowers Amazon climate threshold. *Nature*. 654(8117):114–20

85. Van Nes EH, Holmgren M, Hirota M, Scheffer M. 2012. Response to Comment on "Global Resilience of Tropical Forest and Savanna to Critical Transitions." *Science (1979).* 336(6081):541–541

86. Patterson D, Levin S, Staver AC, Touboul J. 2024. Pattern Formation in Mesic Savannas. *Bull. Math. Biol.* 86(1):3

87. Mathias J-D, Deffuant G, Brias A. 2024. From tipping point to tipping set: Extending the concept of regime shift to uncertain dynamics for real-world applications. *Ecol. Modell.* 496:110801

88. Bultan S, Moustakis Y, Bathiany S, Boers N, Ganzenmüller R, et al. 2025. Amazon forest faces severe decline under the dual pressures of anthropogenic climate change and land-use change. *Proceedings of the National Academy of Sciences*. 122(50):

89. Oliveras I, Malhi Y. 2016. Many shades of green: the dynamic tropical forest–savannah transition zones. *Philosophical Transactions of the Royal Society B: Biological Sciences*. 371(1703):20150308

90. Aleman JC, Fayolle A, Favier C, Staver AC, Dexter KG, et al. 2020. Floristic evidence for alternative biome states in tropical Africa. *Proceedings of the National Academy of Sciences*. 117(45):28183–90

91. Boulton CA, Lenton TM, Boers N. 2022. Pronounced loss of Amazon rainforest resilience since the early 2000s. *Nat. Clim. Chang.* 12(3):271–78

92. Flores BM, Holmgren M. 2021. White-Sand Savannas Expand at the Core of the Amazon After Forest Wildfires. *Ecosystems*. 24(7):1624–37

93. Metz T, Farwig N, Dormann CF, Schaefer HM, Guevara-Andino JE, et al. 2026. Biodiversity resilience in a tropical rainforest. *Nature*. 652(8112):1232–39

94. Lapola DM, Pinho P, Barlow J, Aragão LEOC, Berenguer E, et al. 2023. The drivers and impacts of Amazon forest degradation. *Science (1979).* 379(6630):

95. van Nes EH, Hirota M, Holmgren M, Scheffer M. 2014. Tipping points in tropical tree cover: linking theory to data. *Glob. Chang. Biol.* 20(3):1016–21

96. Stommel H. 1961. Thermohaline Convection with Two Stable Regimes of Flow. *Tellus*. 13(2):224–30

97. Levermann A, Winkelmann R. 2016. A simple equation for the melt elevation feedback of ice sheets. *Cryosphere*. 10(4):1799–1807

98. Rocha JC, Peterson G, Bodin Ö, Levin S. 2018. Cascading regime shifts within and across scales. *Science (1979).* 362(6421):1379–83

99. Wunderling N, von der Heydt AS, Aksenov Y, Barker S, Bastiaansen R, et al. 2024. Climate tipping point interactions and cascades: a review. *Earth System Dynamics*. 15(1):41–74

100. Scheffer M, Carpenter S, Foley JA, Folke C, Walker B. 2001. Catastrophic shifts in ecosystems. *Nature*. 413(6856):591–96

101. Klose AK, Karle V, Winkelmann R, Donges JF. 2020. Emergence of cascading dynamics in interacting tipping elements of ecology and climate. *R. Soc. Open Sci.* 7(6):200599

102. Dekker MM, von der Heydt AS, Dijkstra HA. 2018. Cascading transitions in the climate system. *Earth System Dynamics*. 9(4):1243–60

103. Wunderling N, Krönke J, Wohlfarth V, Kohler J, Heitzig J, et al. 2021. Modelling nonlinear dynamics of interacting tipping elements on complex networks: the PyCascades package. *Eur. Phys. J. Spec. Top.* 230(14–15):3163–76

104. Sinet S, von der Heydt AS, Dijkstra HA. 2023. AMOC Stabilization Under the Interaction With Tipping Polar Ice Sheets. *Geophys. Res. Lett.* 50(2):

105. Klose AK, Donges JF, Feudel U, Winkelmann R. 2024. Rate-induced tipping cascades arising from interactions between the Greenland Ice Sheet and the Atlantic Meridional Overturning Circulation. *Earth System Dynamics*. 15(3):635–52

106. Rahmstorf S, Crucifix M, Ganopolski A, Goosse H, Kamenkovich I, et al. 2005. Thermohaline circulation hysteresis: A model intercomparison. *Geophys. Res. Lett.* 32(23):

107. Garbe J, Albrecht T, Levermann A, Donges JF, Winkelmann R. 2020. The hysteresis of the Antarctic Ice Sheet. *Nature*. 585(7826):538–44

108. van Westen RM, Kliphuis M, Dijkstra HA. 2024. Physics-based early warning signal shows that AMOC is on tipping course. *Sci. Adv.* 10(6):

109. Parry IM, Ritchie PDL, Cox PM. 2022. Evidence of localised Amazon rainforest dieback in CMIP6 models. *Earth System Dynamics*. 13(4):1667–75

110. Ben-Yami M, Good P, Jackson LC, Crucifix M, Hu A, et al. 2024. Impacts of AMOC Collapse on Monsoon Rainfall: A Multi-Model Comparison. *Earths Future*. 12(9):

111. Sinet S, von der Heydt AnnaS, Dijkstra HA. 2025. Meltwater from West Antarctic ice sheet tipping affects AMOC resilience. *Sci. Adv.* 11(46):

112. Vinković D, Kirman A. 2006. A physical analogue of the Schelling model. *Proceedings of the National Academy of Sciences*. 103(51):19261–65

113. Rothstein R. 2017. *The color of law: A forgotten history of how our government segregated America*, Vol. 500. New York: Liveright Publishing Corporation

114. Grodzins M. 1957. Metropolitan Segregation. *Scientific American Magazine*. 197(4):33–41

115. Menendian S, Gambhir S, Gailes A. 2025. The Roots of Structural Racism Project. Othering & Belonging Institute. https://belonging.berkeley.edu/roots-structural-racism-project-2025-update

116. Zhang J. 2004. Residential segregation in an all-integrationist world. *J. Econ. Behav. Organ.* 54(4):533–50

117. Sakoda JM. 1949. *Minidoka: an analysis of changing patterns of social interaction*. University of California, Berkeley

118. Sakoda JM. 1971. The checkerboard model of social interaction. *J. Math. Sociol.* 1(1):119–32

119. Hegselmann R. 2017. Thomas C. Schelling and James M. Sakoda: The Intellectual, Technical, and Social History of a Model. *Journal of Artificial Societies and Social Simulation*. 20(3):

120. Schelling TC. 1971. Dynamic models of segregation. *J. Math. Sociol.* 1(2):143–86

121. Blair P. 2023. Beyond Racial Attitudes: The Role of Outside Options in the Dynamics of White Flight. Cambridge, MA

122. Card D, Mas A, Rothstein J. 2008. Tipping and the Dynamics of Segregation. *Quarterly Journal of Economics*. 123(1):177–218

123. Goering JM. 1978. Neighborhood Tipping and Racial Transition: A Review of Social Science Evidence. *J. Am. Inst. Plann.* 44(1):68–78

124. Logan TD, Parman JM. 2025. Racial Residential Segregation in the United States. *J. Econ. Lit.* 63(3):964–1010

125. Bicchieri C. 2005. *The Grammar of Society*. Cambridge University Press

126. Young HP. 2015. The Evolution of Social Norms. *Annu. Rev. Econom.* 7:359–87

127. Constantino SM, Sparkman G, Kraft-Todd GT, Bicchieri C, Centola D, et al. 2022. Scaling Up Change: A Critical Review and Practical Guide to Harnessing Social Norms for Climate Action. *Psychological Science in the Public Interest*. 23(2):50–97

128. Centola D, Becker J, Brackbill D, Baronchelli A. 2018. Experimental evidence for tipping points in social convention. *Science (1979).* 360(6393):1116–19

129. Ehret S, Constantino SM, Weber EU, Efferson C, Vogt S. 2022. Group identities can undermine social tipping after intervention. *Nat. Hum. Behav.* 6(12):1669–79

130. Efferson C, Vogt S, Fehr E. 2019. The promise and the peril of using social influence to reverse harmful traditions. *Nat. Hum. Behav.* 4(1):55–68

131. Centola D. 2021. *Change: How to make big things happen*. Hachette UK

132. Bollinger B, Gillingham K. 2012. Peer Effects in the Diffusion of Solar Photovoltaic Panels. *Marketing Science*. 31(6):900–912

133. Rode J, Weber A. 2016. Does localized imitation drive technology adoption? A case study on rooftop photovoltaic systems in Germany. *J. Environ. Econ. Manage.* 78:38–48

134. Rogers EM. 1962. *Diffusion of Innovations*. Free Press of Glencoe. 372 pp. 1st ed.

135. Bass FM. 1969. A New Product Growth for Model Consumer Durables. *Manage. Sci.* 15(5):215–27

136. Granovetter M. 1978. Threshold Models of Collective Behavior. *American Journal of Sociology*. 83(6):1420–43

137. Farmer JD, Hepburn C, Ives MC, Hale T, Wetzer T, et al. 2019. Sensitive intervention points in the post-carbon transition. *Science (1979).* 364(6436):132–34

138. Otto IM, Donges JF, Cremades R, Bhowmik A, Hewitt RJ, et al. 2020. Social tipping dynamics for stabilizing Earth's climate by 2050. *Proc. Natl. Acad. Sci. U. S. A.* 117(5):2354–65

139. Tindall DB, Howe AC, Mauboulès C. 2021. Tangled Roots: Personal Networks and the Participation of Individuals in an Anti-environmentalism Countermovement. *Sociological Perspectives*. 64(1):5–36

140. Pereira LM, Gianelli I, Achieng T, Amon D, Archibald S, et al. 2024. Equity and justice should underpin the discourse on tipping points. *Earth System Dynamics*. 15(2):341–66

141. Azariadis C, Stachurski J. 2005. *Poverty traps*, Vol. 1. 295-384. pp.

142. Bowles S, DSN, & HK (Eds. ). 2011. *Poverty traps*. Princeton University Press

143. Barrett CB, Garg T, McBride L. 2016. Well-Being Dynamics and Poverty Traps. *Annu. Rev. Resour. Economics*. 8(1):303–27

144. Allison HE, Hobbs RJ. 2004. Resilience, Adaptive Capacity, and the "Lock-in Trap" of the Western Australian Agricultural Region. *Ecology and Society*. 9(3):

145. Carpenter SR, Brock WA. 2008. Adaptive capacity and traps. *Ecology & Society*. 13(2)(40):

146. Lade SJ, Haider LJ, Engström G, Schlüter M. 2017. Resilience offers escape from trapped thinking on poverty alleviation. *Sci. Adv.* 3(5):

147. Gunderson LH, Holling CS. 2002. *Panarchy: Understanding transformations in human and natural systems*. Washington, D.C.: Island Press. 507 pp.

148. Westley F, Olsson P, Folke C, Homer-Dixon T, Vredenburg H, et al. 2011. Tipping Toward Sustainability: Emerging Pathways of Transformation. *Ambio*, pp. 1–21

149. STENECK RS, HUGHES TP, CINNER JE, ADGER WN, ARNOLD SN, et al. 2011. Creation of a Gilded Trap by the High Economic Value of the Maine Lobster Fishery. *Conservation Biology*. 25(5):904–12

150. Boonstra WJ, de Boer FW. 2014. The Historical Dynamics of Social–Ecological Traps. *Ambio*. 43(3):260–74

151. Haider LJ, Boonstra WJ, Peterson GD, Schlüter M. 2018. Traps and Sustainable Development in Rural Areas: A Review. *World Dev.* 101:311–21

152. Azariadis C. 2011. Chapter 1. The Theory of Poverty Traps. What Have We Learned? In *Poverty Traps*, pp. 17–40. Princeton University Press

153. Mehlum H, Moene K, Torvik R. 2011. Chapter 3. Parasites. In *Poverty Traps*, pp. 79–94. Princeton University Press

154. Bowles S. 2011. Chapter 5. Institutional Poverty Traps. In *Poverty Traps*, pp. 116–38. Princeton University Press

155. Durlauf SN. 2011. Chapter 6. Groups, Social Influences, and Inequality. In *Poverty Traps*, pp. 141–75. Princeton University Press

156. Barrett CB, Swallow BM. 2006. Fractal poverty traps. *World Dev.* 34(1):1–15

157. Ngonghala CN, De Leo GA, Pascual MM, Keenan DC, Dobson AP, Bonds MH. 2017. General ecological models for human subsistence, health and poverty. *Nat. Ecol. Evol.* 1(8):1153–59

158. Radosavljevic S, Haider LJ, Lade SJ, Schlüter M. 2021. Implications of poverty traps across levels. *World Dev.* 144:105437

159. Banerjee AV., Duflo Esther. 2012. *Poor economics : a radical rethinking of the way to fight global poverty*. PublicAffairs. 303 pp.

160. Geels FW. 2002. Technological transitions as evolutionary reconfiguration processes: a multi-level perspective and a case-study. *Res. Policy*. 31:1257–74

161. Geels FW, Schot J. 2007. Typology of sociotechnical transition pathways. *Res. Policy*. 36(3):399–417

162. Geels FW. 2019. Socio-technical transitions to sustainability: a review of criticisms and elaborations of the Multi-Level Perspective. *Curr. Opin. Environ. Sustain.* 39:187–201

163. Kauffman S, Levin S. 1987. Towards a general theory of adaptive walks on rugged landscapes. *J. Theor. Biol.* 128(1):11–45

164. Rotmans J, Loorbach D. 2009. Complexity and Transition Management. *J. Ind. Ecol.* 13(2):184–96

165. Grin J, Rotmans J, Schot J. 2010. *Transitions to Sustainable Development*. Routledge

166. Rotmans J, Kemp R, van Asselt M. 2001. More evolution than revolution: transition management in public policy. *Foresight*. 3(1):15–31

167. de Haan J (Hans), Rotmans J. 2011. Patterns in transitions: Understanding complex chains of change. *Technol. Forecast. Soc. Change*. 78(1):90–102

168. Köhler J, de Haan F, Holtz G, Kubeczko K, Moallemi E, et al. 2018. Modelling Sustainability Transitions: An Assessment of Approaches and Challenges. *Journal of Artificial Societies and Social Simulation*. 21(1):

169. Zeppini P, Frenken K, Kupers R. 2014. Thresholds models of technological transitions. *Environ. Innov. Soc. Transit.* 11:54–70

170. Bergman N, Haxeltine A, Whitmarsh L, Köhler J, Schilperoord M, Rotmans J. 2008. Modelling socio-technical transition patterns and pathways. *J. Artif. Soc. Soc. Simul.* 11:3(7):

171. Smith A, Raven R. 2012. What is protective space? Reconsidering niches in transitions to sustainability. *Res. Policy*. 41(6):1025–36

172. Kern F, Verhees B, Raven R, Smith A. 2015. Empowering sustainable niches: Comparing UK and Dutch offshore wind developments. *Technol. Forecast. Soc. Change*. 100:344–55

173. Kern F, Smith A. 2008. Restructuring energy systems for sustainability? Energy transition policy in the Netherlands. *Energy Policy*. 36(11):4093–4103

174. Geels FW. 2014. Regime Resistance against Low-Carbon Transitions: Introducing Politics and Power into the Multi-Level Perspective. *Theory Cult. Soc.* 31(5):21–40

175. Avelino F. 2021. Theories of power and social change. Power contestations and their implications for research on social change and innovation. *Journal of Political Power*. 14(3):425–48

176. Avelino F, Rotmans J. 2009. Power in Transition: An Interdisciplinary Framework to Study Power in Relation to Structural Change. *European Journal of Social Theory*. 12(4):543–69

177. Avelino F, Wittmayer JM. 2016. Shifting Power Relations in Sustainability Transitions: A Multi-actor Perspective. *Journal of Environmental Policy & Planning*. 18(5):628–49

178. Moore M-L, Milkoreit M. 2020. Imagination and transformations to sustainable and just futures. *Elementa: Science of the Anthropocene*. 8(1):

179. Clark WC, Harley AG. 2020. Sustainability Science: Toward a Synthesis. *Annu. Rev. Environ. Resour.* 45:331–86

180. Folke C, Carpenter SR, Walker B, Scheffer M, Chapin T, Rockström J. 2010. Resilience thinking: Integrating resilience, adaptability and transformability. *Ecology and Society*. 15(4):

181. Olsson P, Folke C, Hahn T. 2004. Social-Ecological Transformation for Ecosystem Management: The Development of Adaptive Co-Management of a Wetland Landscape in Southern Sweden. *Ecology & Society*. 9(4):

182. Olsson P, Gunderson LH, Carpenter SR, Ryan P, Lebel L, et al. 2006. Shooting the rapids: Navigating transitions to adaptive governance of social-ecological systems. *Ecology and Society*. 11(1):

183. Pereira L, Frantzeskaki N, Hebinck A, Charli-Joseph L, Drimie S, et al. 2020. Transformative spaces in the making: key lessons from nine cases in the Global South. *Sustain. Sci.* 15(1):161–78

184. Olsson P, Moore M-L. 2024. Transformations, Agency and Positive Tipping Points: A Resilience-Based Approach. , pp. 59–77

185. O´Brien K, Sygna L. 2013. Responding to climate change: The three spheres of transformation

186. Schill C, Anderies JM, Lindahl T, Folke C, Polasky S, et al. 2019. A more dynamic understanding of human behaviour for the Anthropocene. *Nat. Sustain.* 2(December):1075–82

187. Chambers JM, Wyborn C, Klenk NL, Ryan M, Serban A, et al. 2022. Co-productive agility and four collaborative pathways to sustainability transformations. *Global Environmental Change*. 72:

188. Whyte KP. 2018. Indigenous science (fiction) for the Anthropocene: Ancestral dystopias and fantasies of climate change crises. *Environ. Plan. E Nat. Space*. 1(1–2):224–42

189. Bennett EM, Solan M, Biggs R, McPhearson T, Norström A V, et al. 2016. Bright spots: seeds of a good Anthropocene. *Frontiers in Ecology*. 14(8):441–48

190. Argyris C, Schön DA. 1978. *Organizational learning: a theory of action perspective*. Reading, Massachussets: Addison-Wesley Pub. Co.

191. Meadows DH. 1999. Leverage Points: Places to Intervene in a System

192. Scoones I, Stirling A, Abrol D, Atela J, Charli-joseph L, et al. 2020. Transformations to sustainability : combining structural , systemic and enabling approaches. *Curr. Opin. Environ. Sustain.* 42:65–75

193. Krueger EH, Constantino SM, Centeno MA, Elmqvist T, Weber EU, Levin SA. 2022. Governing sustainable transformations of urban social- ecological-technological systems. *npj Urban Sustainability*. 2:10:1–12

194. Olsson P, Folke C, Hughes TP. 2008. Navigating the transition to ecosystem-based management of the Great Barrier Reef, Australia. *Proceedings of the National Academy of Sciences*. 105(28):9489–94

195. Hillebrand H, Donohue I, Harpole WS, Hodapp D, Kucera M, et al. 2020. Thresholds for ecological responses to global change do not emerge from empirical data. *Nat. Ecol. Evol.* 4(11):1502–9

196. FRANCIS RC, HARE SR. 1994. Decadal-scale regime shifts in the large marine ecosystems of the North-east Pacific: a case for historical science. *Fish. Oceanogr.* 3(4):279–91

197. Mantua NJ, Hare SR. 2002. The Pacific Decadal Oscillation. *J. Oceanogr.* 58(1):35–44

198. Hare SR, Mantua NJ. 2000. Empirical evidence for North Pacific regime shifts in 1977 and 1989. *Prog. Oceanogr.* 47(2–4):103–45

199. Scheffer M, Bascompte J, Brock WA, Brovkin V, Carpenter SR, et al. 2009. Early-warning signals for critical transitions. *Nature*. 461(7260):53–59

200. Dakos V, Scheffer M, van Nes EH, Brovkin V, Petoukhov V, Held H. 2008. Slowing down as an early warning signal for abrupt climate change. *Proceedings of the National Academy of Sciences*. 105(38):14308–12

201. Abson DJ, Fischer J, Leventon J, Newig J, Schomerus T, et al. 2017. Leverage points for sustainability transformation. *Ambio*. 46(1):30–39

202. Scheffer M, Carpenter SR. 2003. Catastrophic regime shifts in ecosystems: linking theory to observation. *Trends Ecol. Evol.* 18(12):648–56